\documentclass[twocolumn,twocolappendix,trackchanges]{aastex63}
\usepackage{amsmath}
\usepackage{multirow}

\newcommand{\pair}{\ensuremath{e^\pm}}

\newcommand{\msun}{\ensuremath{\,\rm{M}_\odot}}

\newcommand{\nickel}{\ensuremath{^{56}\mathrm{Ni}}}
\newcommand{\ergs}{\ensuremath{\rm{ergs}}}

\newcommand{\maxbh}{\ensuremath{\rm{M}_{\rm{BH,max}}}}
\newcommand{\mheint}{\ensuremath{\rm{M_{He,init}}}}

\newcommand{\mbh}{\ensuremath{\rm{M_{\rm BH}}}}
\newcommand{\crate}{\ensuremath{^{12}\rm{C}\left(\alpha,\gamma\right)^{16}\!\rm{O}}}

\newcommand{\kevbarns}{\ensuremath{\rm{\,\kev\, barns}}}

\newcommand{\csigma}{\ensuremath{\sigma_{\rm{C12}}}}

\newcommand{\code}[1]{\texttt{#1}}

\newcommand{\MESA}{\code{MESA}}
\newcommand{\STARLIB}{\code{STARLIB}}
\newcommand{\REACLIB}{\code{REACLIB}}
\newcommand{\NACRE}{\code{NACRE}}

\newcommand{\kms}{{\mathrm{km\ s^{-1}}}}

\newcommand{\kev}{\mathrm{keV}}
\newcommand{\gk}{\ensuremath{\,\rm{GK}}}

\DeclareRobustCommand{\Eqref}[1]{Equation ~\ref{#1}}

\DeclareRobustCommand{\eqref}[1]{\Eqref{#1}}

\newcommand{\ts}{\textsuperscript}

\newcommand{\nuclei}[2]{\ensuremath{\mathrm{^{#1}#2}}}

\newcommand{\helium}[1][4]{\nuclei{#1}{He}}

\newcommand{\carbon}[1][12]{\nuclei{#1}{C}}

\newcommand{\oxygen}[1][16]{\nuclei{#1}{O}}

\newcommand{\neon}[1][20]{\nuclei{#1}{Ne}}
\newcommand{\sodium}[1][23]{\nuclei{#1}{Na}}
\newcommand{\magnesium}[1][24]{\nuclei{#1}{Mg}}
\newcommand{\silicon}[1][28]{\nuclei{#1}{Si}}

\begin{document}

\title{Constraints from gravitational wave detections of binary black hole mergers on the \crate{} rate}


\author[0000-0003-3441-7624]{R.~Farmer}
\email{r.j.farmer@uva.nl}
\affiliation{Anton Pannekoek Institute for Astronomy and GRAPPA, University of Amsterdam, 
NL-1090 GE Amsterdam, The Netherlands}
\affiliation{Center for Astrophysics | Harvard \& Smithsonian, 60 Garden Street, Cambridge, MA 02138, USA}

\author[0000-0002-6718-9472]{M.~Renzo}
\affiliation{Center for Computational Astrophysics, Flatiron Institute, New York, NY 10010, USA}

\author[0000-0001-9336-2825]{S.~E.~de~Mink}
\affiliation{Center for Astrophysics | Harvard \& Smithsonian, 60 Garden Street, Cambridge, 
MA 02138, USA}
\affiliation{Anton Pannekoek Institute for Astronomy and GRAPPA, University of Amsterdam, 
NL-1090 GE Amsterdam, The Netherlands}

\author[0000-0002-1980-5293]{M.~Fishbach}
\affiliation{Department of Astronomy and Astrophysics, University of Chicago, Chicago, IL 60637, USA}

\author[0000-0001-7969-1569]{S.~Justham}
\affiliation{School of Astronomy \& Space Science, University of the Chinese Academy of Sciences, 
Beijing 100012, China}
\affiliation{National Astronomical Observatories, Chinese Academy of Sciences, Beijing 100012, China}
\affiliation{Anton Pannekoek Institute for Astronomy and GRAPPA, University of Amsterdam, NL-1090 
GE Amsterdam, The Netherlands}

\date{\today}

\begin{abstract}

Gravitational wave detections are starting to allow us to probe the physical processes in the evolution of very massive stars through the imprints they leave on their final remnants.  
Stellar evolution theory predicts the existence of a gap in the black hole mass distribution at high mass due to the effects of pair-instability.  
Previously, we showed that the location of the gap is robust against model uncertainties, but it does depend sensitively on the uncertain \crate{} rate. This rate is of great astrophysical significance and governs the production of oxygen at the expense of carbon.
We use the open source \MESA{} stellar evolution code to evolve massive helium stars to probe the location of the mass gap. 
We find that the maximum black hole mass below the gap varies between $40\msun$ to $90\msun$, depending on the strength of the uncertain \crate{} reaction rate. 
With the first ten gravitational-wave detections of black holes, we constrain the astrophysical S-factor for \crate{}, at $300\kev$, to $S_{300}>175\kevbarns$ at 68\% confidence. With $\mathcal{O}(50)$ detected binary black hole mergers, we expect to constrain the S-factor to within $\pm10$--$30\kevbarns$. 
We also highlight a role for independent constraints from electromagnetic transient surveys. The unambiguous detection of pulsational pair instability supernovae would imply that $S_{300}>79\kevbarns$.
Degeneracies with other model uncertainties need to be investigated further, but probing nuclear stellar astrophysics poses a promising science case for the future gravitational wave detectors.

\end{abstract}


\section{Introduction}\label{sec:introduction}

The \crate{} reaction is one of the most important nuclear reaction rates \citep{burbidge57}
in the evolution of stars yet also one of the most uncertain \citep{holt19}. 
Reducing the uncertainty on this rate has been dubbed 
``the holy grail of nuclear astrophysics" \citep{deboer17,bemmerer18}.
It plays a key role in governing the evolution and composition of stars beyond the main
sequence, from the C/O ratio in white dwarfs \citep{salaris97,straniero03,fields16},
whether a
star will form a neutron star or a black hole \citep{brown01,woosley:02,heger02b,tur07,west13,sukhbold20},
and the amount of  \carbon{} and \oxygen{} in the Universe \citep{boothroyd88,weaver93b,thielemann96}

Thus improving our understanding of this key rate is of critical importance
to stellar astrophysics. The difficulty in measuring the rate
occurs due to the negligible cross section of the reaction at temperatures
relevant for helium burning in stars \citep{an15,an16}. Thus nuclear experiments 
can only provide data for much higher energies (i.e. temperatures) 
from which we extrapolate down to astrophysically relevant energies. However, 
the cross-section has a complex energy dependence and thus
is not easily extrapolated to lower temperatures \citep{deboer17,friscic19}. 
Recent lab measurements, with high beam luminosities, 
though indirect studies of the excited states of \oxygen{}, and
improved theoretical modeling of the \crate{} rate 
have begun to reduce the uncertainty on the rate \citep{hammer02,an16,hammache16,deboer17,shen20}. 
New experiments will soon be better able to probe the \crate{} reaction rate at astrophysically
relevant temperatures \citep{holt18,friscic19}.

Astrophysical studies using white dwarfs have attempted to place constraints on the 
\crate{} reaction rate, using astroseismology of white dwarfs \citep{metcalfe01,metcalfe02,metcalfe03}.
However, these measurements are sensitive to other physics choices, like 
semiconvection and convective overshoot mixing \citep{straniero03}, that are
poorly constrained. Thus a cleaner signal is needed to provide a more robust estimate from stellar astrophysical sources. 

Merging black holes detected by LIGO/Virgo \citep{ligo15a,virgo15} can provide such a signal, 
via the location of the
pair-instability mass gap \citep{takahashi18,farmer:19}.
A gap is predicted to form in the mass distribution of black holes, due to
pair-instability supernovae (PISN) completely disrupting massive stars, 
leaving behind no remnant
\citep{fowler:64, barkat:67, woosley:17}. The lower edge of the gap
is set by mass loss experienced by a star during pulsational pair instability
supernovae \citep{rakavy:67,fraley:68, woosley:02}. These objects undergo
multiple mass-loss phases before collapsing into a black hole 
\citep{woosley:02,chen:14,yoshida:16, woosley:17,marchant:19,farmer:19}.

In \citet{farmer:19} we evolved hydrogen-free helium cores 
and found that the lower edge of the
PSIN black hole mass gap was robust to changes in the metallicity and other uncertain physical processes,
e.g wind mass loss and chemical mixing.
Over the range of metallicities considered the maximum black hole mass
decreased by ~3\msun. We also showed that the
choices for many other uncertain physical processes inside stars do not greatly affect the location of the PISN mass gap.

The existence of a gap in the mass distribution of merging binary black holes (BBH) would
provide strong constraints on their progenitors, and hence of the post main-sequence evolution
of stars, which includes the effect of \crate{} on a star's evolution \citep{takahashi18,sukhbold20}.
The existence of a gap in the mass distribution can also be used as a 
``standardizable sirens'' for cosmology and used to place
constraints on the Hubble constant \citep{schutz:86,holz15,farr19}.

Here we investigate how
the maximum black hole mass below the PISN mass gap is sensitive to the
\crate{} nuclear reaction rate and thus can be used to place constraints on the
reaction rate. In Section \ref{sec:method} we discuss our methodology. In Section \ref{sec:results}
we describe the star's evolution before pulsations begin, and how this is altered by the \crate{} reaction rate.
In Section \ref{sec:edgegap} we show how the maximum black hole mass below the gap is affected by the
nuclear redaction rates and place constraints on the \crate{} reaction rate in
Section \ref{sec:rates}. 
In Section \ref{sec:future} we discuss how these results
will improve with future gravitational wave detections. 
In Section \ref{sec:bhimf} we discuss potentially other observables
that can be used to constrain the \crate{}.
Finally, in Sections \ref{sec:discuss} \& \ref{sec:conclusion},
we discuss and summarize our results.

\section{Method}\label{sec:method}

There are many channels for the formation of a source detectable by 
ground-based gravitational-wave detectors. 
We consider here the case where the progenitors of the merging black holes have come from an isolated
binary system. There are multiple stellar pathways for this to produce a successful 
binary black hole merger,
including common-envelope evolution \citep{tutukov93,dominik:12,belczynski:16nat}, 
chemically homogeneous evolution 
\citep{demink16,marchant:16,mandel:16b}, or stars which interact in dynamic environments 
\citep{Kulkarni93,Portegies00,gerosa19}
In each case, we expect the stars to lose their hydrogen envelopes 
after the end of the star's main sequence, leaving
behind the helium core of the star.

We use the \MESA{} stellar evolution code, version 11701 
\citep{paxton:11,paxton:13,paxton:15,paxton:18,paxton:19},
to follow through the various stages of nuclear burning
inside these helium cores until they either collapse to
form a black hole or explode as a PISN. 
We follow the evolution of helium cores with initial masses between $\mheint=30\msun$ and
$\mheint=200\msun$ in steps of $1\msun$, at a metallicity of $\rm{Z}=10^{-5}$.
We use the default model choices
from \citet{farmer:19} for setting all other \MESA{} input parameters.
See Appendix \ref{sec:appen_mesa} for further details of our usage of \MESA{}, and
our input files with all the parameters we set can be found at 
\url{https://zenodo.org/record/3559859}.

After a star has formed its helium core, it begins burning helium in its central region,
converting \helium{} into
\carbon{} and then \oxygen{}. The final ratio of the mass fractions of \carbon/\oxygen{} depends
on the relative strengths of the $3\alpha$ reaction rate, which produces \carbon{}, and
\crate{} reaction rate, which converts the \carbon{} into \oxygen{}.
We define the end of core helium burning to occur when  the central mass fraction of \helium{} drops below
$10^{-4}$. The core is now dominated by \carbon{} and \oxygen{}, with only trace mass fractions of
other nuclei. This core then begins a phase of contraction, and
thermal neutrino losses begin to dominate the total energy loss
from the star \citep{fraley:68,heger:03,heger05,farmer16}.

As the core contracts the central density and temperature increases which, for sufficiently massive cores,
causes the core to begin producing copious amounts of electron-positron pairs (\pair).
The production of the \pair{} removes photons which were providing pressure support,
softening the equation of state in the core, and causes the core to contract further.
We then follow the dynamical collapse of the star, which can be halted by the
ignition of oxygen leading to either a PPISN or a PISN. 
We follow the core as it contracts and bounces \citep{marchant:19}, generating shock waves
that we follow through the star until they reach the outer layers of the star.
These shocks then cause mass loss from the star to occur, as material becomes unbound.
In this case we find that the star can eject
between $0.1\msun$ and $\sim20\msun$ of material in a pulsational mass loss episode \citep{woosley:17,farmer:19,renzo20csm}.
Stars at the boundary between core collapse and PPISN may generate weak pulses, 
due to only a small amount of material becoming dynamicaly unstable,
and therefore do not drive any appreciable mass loss \citep{woosley:17}. 
We use the term PPISN only for an event which ejects mass \citep{renzo20csm}.
PISN are stars for which the energy liberated by the thermonuclear explosion of oxygen 
(and carbon) exceeds the total binding energy, resulting in total disruption after only one mass loss episode.

As a star evolves into the pair instability region we switch to using \MESA's Riemen contact solver,
HLLC, \citep{Toro1994,paxton:18}, to follow the hydrodynamical evolution of each pulse.
This switch occurs when the volumetric pressure-weighted 
average adiabatic index $\langle\Gamma_1\rangle - 4/3 < 0.01$,  which occurs slightly before the star enters the
pair instability region. The adiabatic index, $\Gamma_1$ is defined as

\begin{equation}
\label{eq:gamma1}
\Gamma_{1}=\frac{\mathrm{d} \ln P}{\mathrm{d}\ln\rho}\biggr\rvert_{s}
\end{equation}

where $P$, $\rho$ are the local pressure and density and is evaluated at a constant entropy $s$.
We used the continuity
equation to transform the volumetric integral of $\Gamma_1$ into an integral over the mass
domain, thus \citep{stothers:99}:

\begin{equation}
  \label{eq:pulse_criterion}
  \langle\Gamma_1\rangle \equiv \frac{\int \Gamma_1P\,d^3r}{\int P\,d^3r} \equiv \frac{\int \Gamma_1 \frac{P}{\rho}\,dm}{\int
    \frac{P}{\rho}\,dm} 
\end{equation}

We follow the dynamical evolution of the star, until all shocks have reached the surface 
of the star. These shocks may unbind a portion of the outer stellar envelope,
resulting in mass loss \citep{yoshida:16,woosley19,renzo20csm}.
We follow the ejected material until the bound portion of the star relaxes back into hydrostatic
equilibrium, after it has radiated away the energy of the pulse. 
We remove the material that has become unbound from our computational grid 
by generating a new stellar model
with the same entropy and chemical distribution as the remaining bound material.
We evolve this new star assuming hydrostatic equilibrium
until either another pulse occurs or the 
core temperature ($\rm{T_c}$) exceeds $\rm{T_c} >
10^{9.6}\rm{K}$, as the star is approaching core collapse.
At which point we switch back to using the hydrodynamic solver,
We define the final core collapse to occur when any part of the star
begins collapsing with a velocity $v>8000\kms$, so that any pulse that
is in the process of being ejected during core collapse is resolvable.

Stars with core masses above $\mheint \gtrapprox 120\msun$ 
attempt to undergo a PISN, however sufficient energy is released during the pulse 
that the core heats to the point where photo-distintegrations become 
the dominant energy sink. These reactions then
reduce the available energy, which was powering the outward moving shock, 
and prevents the envelope from becoming 
unbound. The star then collapses without significant mass loss. We assume that this
forms a black hole \citep{bond84,woosley:17}.

We define the mass of the black hole formed to be the
mass of the bound material of the star at collapse. Given the uncertain
black hole formation mechanism \citep{fryer99,fryer01,fryer:12}, or weak
shock generation \citep{nadezhin:80,lovegrove:13,fernandez18}, our black holes masses are upper limits.
We take the bound mass not the total mass, as some stars are under going a mass ejection 
from a pulsation at the time of core collapse \citep{renzo20csm}.

\subsection{Nuclear reaction rates}

Nuclear reaction rates play a key role in the evolution and final fate of a star.
However, they are also uncertain and this uncertainty varies as function
of temperature \citep{iliadis_2010_ab,iliadis_2010_aa,longland_2010_aa}.
Varying nuclear reaction rates within their known uncertainties has
been shown to a have large impact on the
structure of a star \citep{hoffman99,iliadis02}.

To sample nuclear reaction rates within their known uncertainties,
we use the \STARLIB{} \citep{sallaska13}
(version 67a) library.
  \STARLIB{} provides both the median reaction rate and uncertainty in that reaction
  as a function of temperature.
  We sample each reaction at a fixed number of standard deviations 
  from the median \citep{evans_2000_aa}.
  We assume that the
  temperature-dependent uncertainty in a reaction follows a log-normal distribution \citet{longland_2010_aa}.

For each reaction rate tested, we create a sampled reaction rate at 60 points 
log-spaced in temperature between,
0.01 and $10\times10^{9}\,\rm{K}$ \citep{fields16,fields18}. 

The rate of a reaction per particle pair is given by:

\begin{multline}
N_A \langle \sigma \nu \rangle = \left(\frac{8}{\pi\mu}\right)^{1/2} \frac{N_A}{\left(k_{B} T\right)^{(3/2)}} \\ 
\int_{0}^{\infty} \sigma\left( E \right) E \exp^{\left( -E/k_{B}T\right)} dE
\end{multline}

where $\mu$ is the reduced mass of the particles, $E=\mu\nu^2/2$ is the center of mass energy, $\nu$ is the average
velocity of the particles,
$N_A$ is Avogadro's number, and $k_B$ is the Boltzmann constant \citep[e.g][]{lippuner17,deboer17,holt19}.
We can factor out the energy dependent cross-section $\sigma\left(E\right)E$, by replacing it with the
astrophysical S-factor:

\begin{equation}\label{eq:sfactor}
S(E) = \sigma\left(E\right)E\exp^{2\pi\eta}
\end{equation}

and 
\begin{equation}
\eta = \sqrt{\frac{\mu}{2E}} Z_1Z_2 \frac{e^2}{\hbar}
\end{equation}

where $\eta$ is the Summerfield parameter, 
$Z_{1,2}$ the proton charge of each particle,
$e$ is the electric charge, $\hbar$ is the reduced Planck's constant. The $\exp^{2\pi\eta}$
term accounts for (approximately) the influence of the Coulomb barrier on the cross-section.
As the S-factor depends on energy, we quote it at the typical energy for a reaction.
For \crate{} the typical energy is $E=300\,\kev$.

\section{Pre-SN carbon burning}\label{sec:results}

\begin{figure}[htp]
  \centering

  \includegraphics[width=\linewidth]{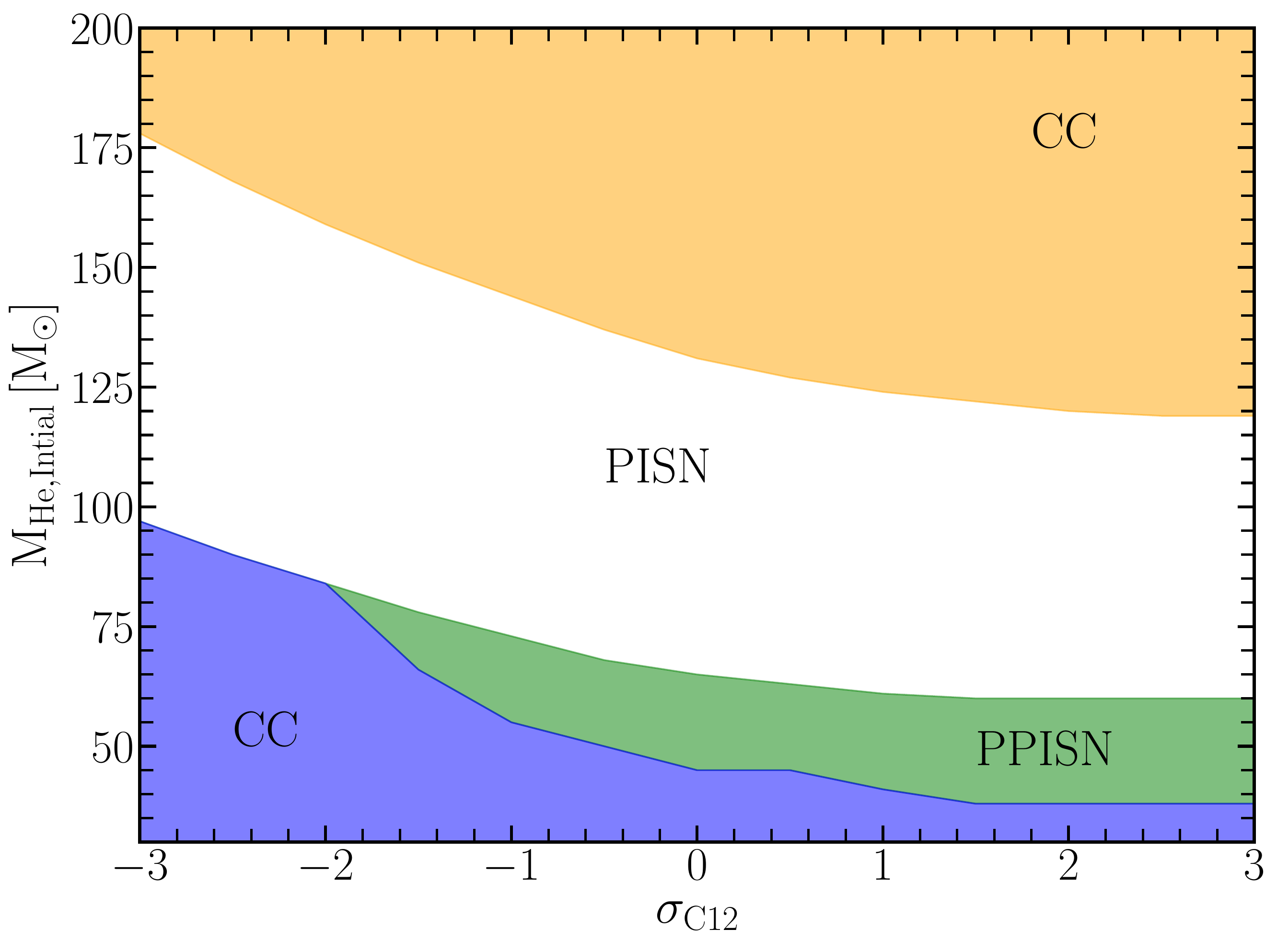}

  \caption{ Final fate of a star as function of the initial helium core mass and \crate{} rate.
  $\csigma$ denotes how far the \crate{} is from the median \STARLIB{} rate, measured in
  standard deviations.
  Blue regions indicate stars which undergo core collapse (CC) below the pair instability supernovae (PISN) mass gap, green regions
  form black holes after a pulsational pair instability supernovae (PPISN), while white regions are completely disrupted in a PISN, and
  models in the orange region form black holes from core collapse for stars above the PISN mass gap. 
  There are 2210 models, in the grid spaced by
  $1\msun$ and $0.5\csigma$.}
    \label{fig:final_state}
\end{figure}

In Fig. \ref{fig:final_state} we show the outcome for our grid of 
2210 evolutionary models as a function of the initial helium core mass 
and the \crate{} reaction. We parameterize the \crate{} in terms of the number of sigmas (\csigma)
from the median \STARLIB{} \crate{} reaction rate:

\begin{equation}
\langle \sigma \nu \rangle =  \langle \sigma \nu \rangle_{\rm{median}} e^{\mu\left(T\right)\csigma}  
\end{equation}

where $\langle \sigma \nu \rangle_{\rm{median}}$ is the median reaction rate provided by \STARLIB{},
and $\mu\left(T\right)$ is the temperature-dependent uncertainty in the reaction (which is assumed to follow a
log-normal distribution). 

For higher initial core mass (for a given \crate{} rate), the final fate of a star 
transitions from core collapse, to PPISN, to PISN, and then to
core collapse again \citep{bond84}.
As the reaction rate increases (i.e, large values of \csigma)  the boundary between the different end fates
shifts to lower initial helium core mass. See Section \ref{sec:bhimf}
for a discussion of the implications of this for the black hole formation and EM transient rate.

\begin{figure}[htp]
  \centering

  \includegraphics[width=\linewidth]{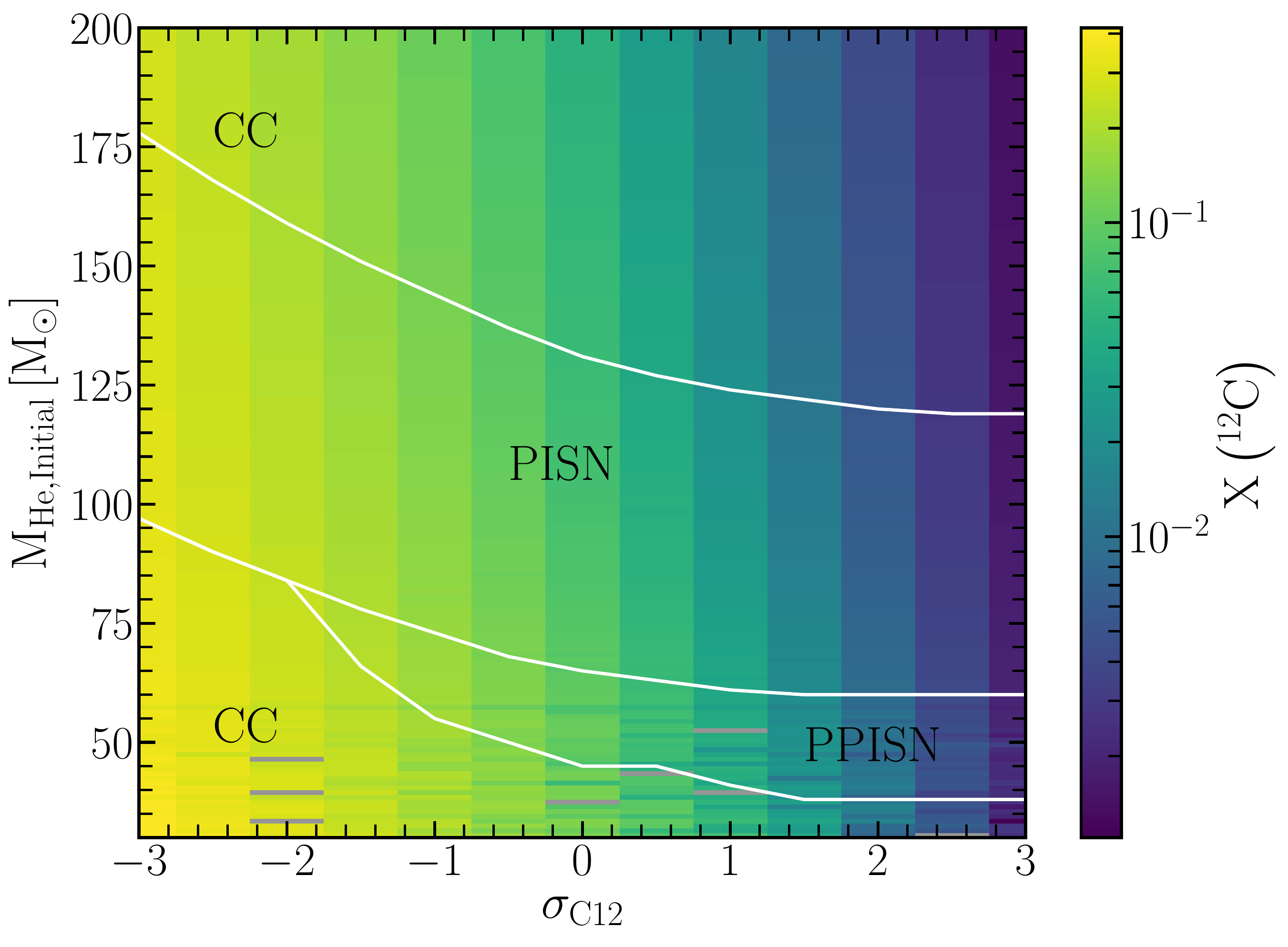}

  \caption{ The mass fraction of \carbon{} in the core after core helium burning, 
  but before carbon burning, for
  all initial masses as a function of \csigma.
   White lines denote the boundaries between the different end fates. Text labels
   denote the final fate of the star.
   Grey boxes denote models that do
   not evolve beyond core helium burning, defined as when the mass fraction of \helium{} at the center
   of the star drops below $10^{-4}$.}
    \label{fig:c12_posthe2d}
\end{figure}

\begin{figure*}[htp]
  \centering

  \includegraphics[width=1.0\textwidth]{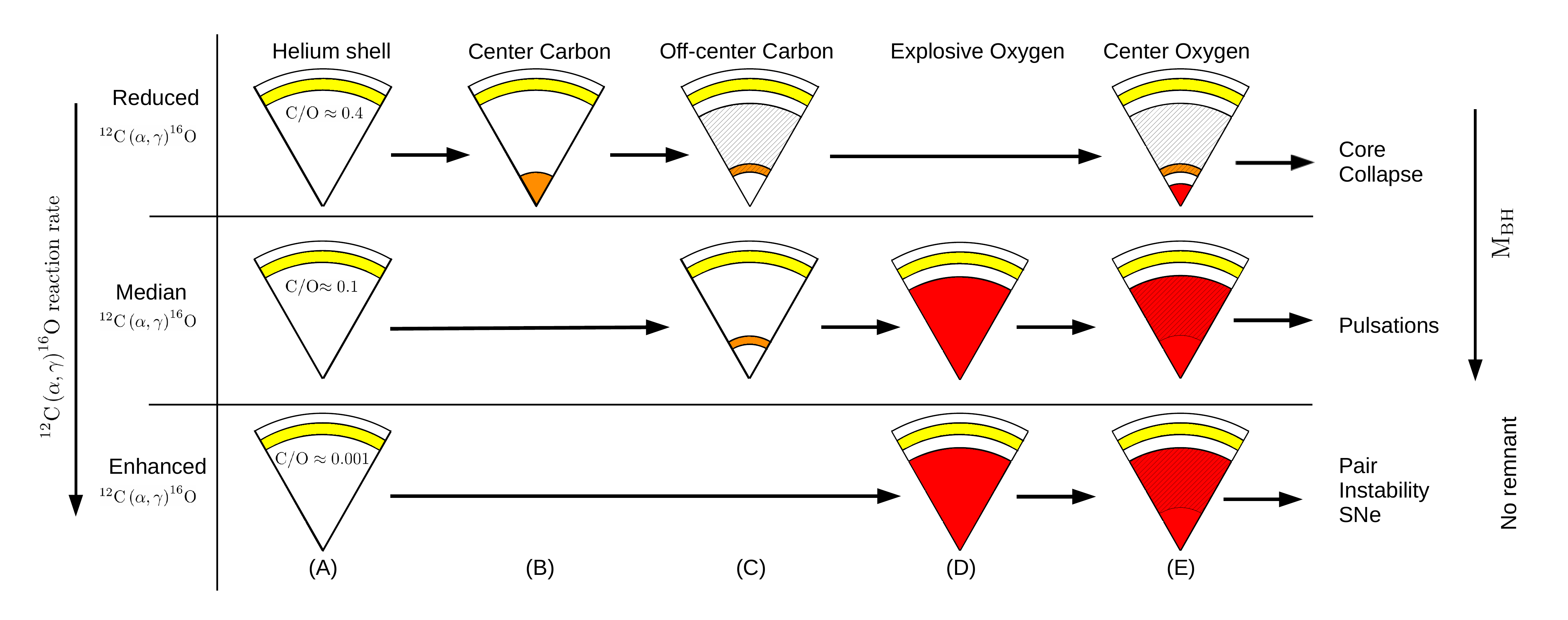}

  \caption{ A schematic of the time progression of the major fuel burning, for a star with $\mheint=60\msun$ as a function of \csigma, at
  reduced ($\csigma=-3$), median ($\csigma=0$), and enhanced ($\csigma=3$) \crate{} reaction rates. Yellow regions denote helium burning,
  orange regions denote carbon burning, and red regions denote oxygen burning. Hatched regions indicate convection mixing regions. Letters match
  those points marked in Fig. \ref{fig:kips}. Also shown is the core's C/O mass ratio at the end of core helium burning.}
    \label{fig:sketch}
\end{figure*}

To understand the reason for these trends, it is insightful to
consider the \carbon{} mass fraction in the core of the stellar models, after core helium burning has finished.
We define the end of core helium burning, when the mass fraction of \helium{} at the center
of the star drops below $10^{-4}$.
Figure \ref{fig:c12_posthe2d} shows that the
\carbon[12] mass fraction in the
core of the stars considered here decreases from $\approx30\%$ to $\approx0.001\%$ as the \crate{} 
rate is increased from $\csigma=-3$ to $\csigma=3$, independent of initial mass. 
This change in the \carbon[12] mass fraction is what drives the changes in the star's later phases of evolution
and thus final fate.

After core helium burning has ceased the core begins contracting, increasing its density and temperature. However,
at the same time, thermal neutrino losses increase which acts to cool the core. 
The next fuel to burn is, \carbon[12]{} via \carbon[12]+\carbon[12]{} to
\neon[22], \sodium[23], and \magnesium[24] \citep{arnett69,farmer15}. 
As the $\carbon[12]+\carbon[12]$ reaction rate
depends on the number density of carbon squared, small changes in the number density of 
\carbon{} can have a large impact on the power generated by the $\carbon[12]+\carbon[12]$ reaction.

In Fig. \ref{fig:sketch} we show a simplified picture of the steps a star takes to its final fate depending on
its \crate{} reaction rate. The top panel shows a star which would undergo core collapse, first by igniting carbon both at the center
and then in a off-center shell. This star then avoids igniting oxygen explosively, instead proceeds through its evolution to core collapse. As the \crate{} increases, the carbon stops igniting at the center and only ignites in a shell (middle panel), before proceeding to ignite oxygen explosively. For the highest \crate{} shown, no carbon is burnt
before oxygen ignites (bottom panel). The C/O ratios shown are defined at the end of core helium burning.

\begin{figure*}[ht]
  \centering
  \includegraphics[width=1.0\textwidth]{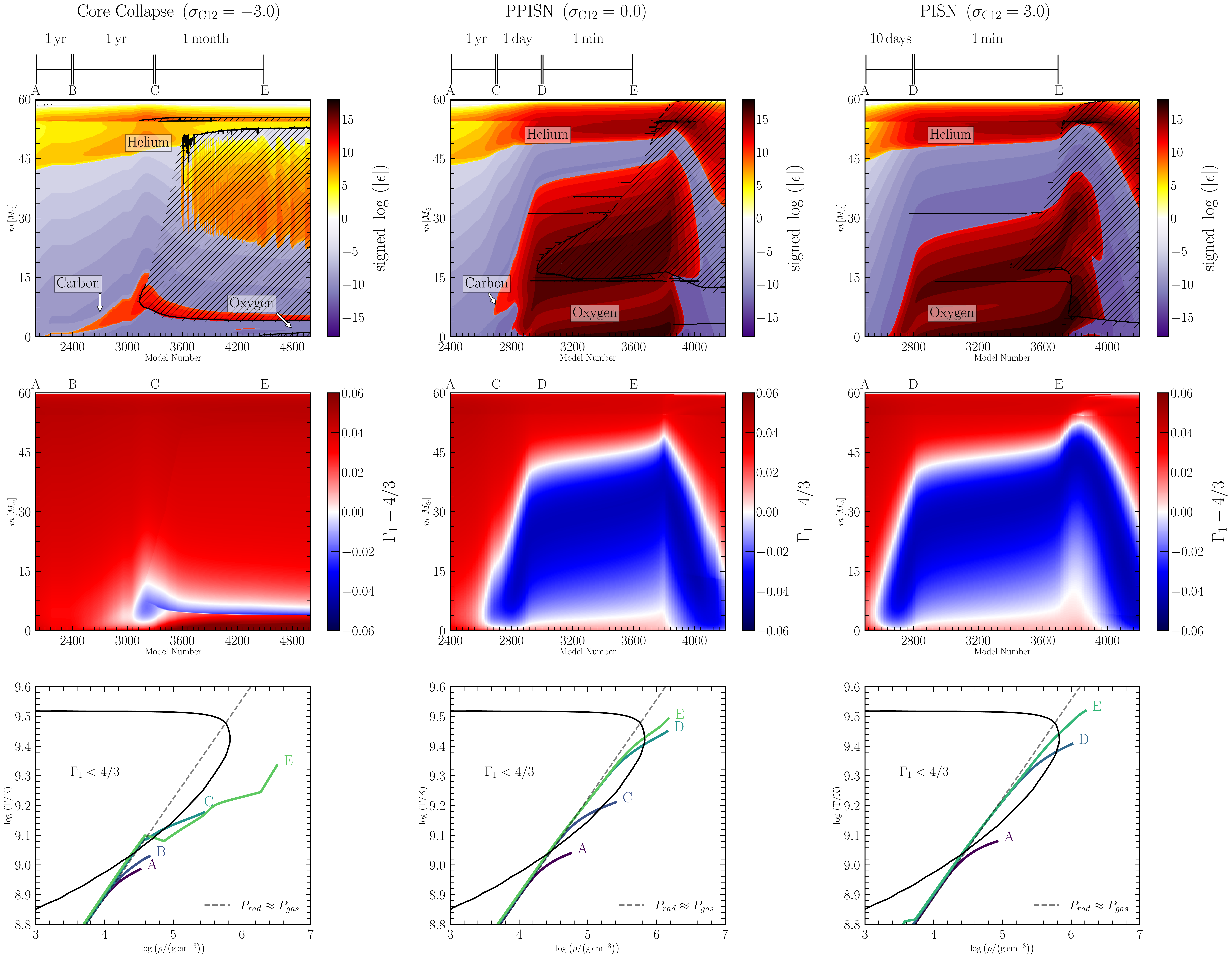}

  \caption{ The time evolution of the internal structure of a star for 
  $\mheint=60\msun$ for different assumptions for the \crate{} reaction rate. The top row shows the signed logarithm of the net specific power, i.e.
  $\rm{sign}\left(\epsilon_{\rm{nuc}}-\epsilon_{\nu}\right)\log_{10}\left(\rm{max}\left(1.0,|\epsilon_{\rm{nuc}}-\epsilon_{\nu}|\right)/\left(\rm{erg\,g^{-1}\, s^{-1}}\right)\right)$, where $\epsilon_{\rm{nuc}}$ 
  is the  specific power from nuclear reactions and $\epsilon_{\nu}$ is the specific power lost via  neutrinos.
   Purple regions denote strong neutrino
   cooling and red regions denote regions of strong nuclear burning. Hatched regions indicate convective mixing regions.
      Text labels state the primary fuel burned in that region. Points marked on the top x axis correspond
  to those marked in Fig. \ref{fig:sketch}, given with approximate timescales.
   The middle row shows the evolution of  $\Gamma_1$ (Equation \ref{eq:gamma1}), regions with $\Gamma_1-4/3 < 0$ are 
   locally unstable.
  The bottom row shows the density-temperature structure of the inner region of the stars at the points marked in the 
  top panel, light colors denote later phases. The dashed line shows
  where the gas pressure is approximately equal to the radiation pressure. 
  The solid black line that encloses $\Gamma_1<4/3$ shows the approximate location of the pair instability region.
}
  \label{fig:kips}
\end{figure*}

Why changing the carbon burning behavior changes the final outcome for a star can be seen in Fig. \ref{fig:kips}.
Here we show the
time evolution of the helium cores, for stars with $\mheint=60\msun$ during carbon burning and up to
the ignition of oxygen, for different \crate{} rates.
The top row shows a Kippenhan diagram of the time evolution of the net nuclear energy 
minus neutrino losses and the mixing regions inside the star. The middle row shows the evolution of
$\Gamma_{1}-4/3$. Regions where  $\Gamma_{1}-4/3 < 0$ are locally unstable.
The bottom row shows the temperature and density structure inside the star at points in time marked
on the top row of Fig. \ref{fig:kips}. The points in time marked, show how the stars are evolving on different
timescales. Timescales vary from a few thermal timescales (left column) to a few dynamical
timescales (middle and right columns).

When the \crate{} rate is small (and thus the \carbon{} mass fraction is $\approx30\%$, with the rest of the core 
being made of \oxygen{}),
\carbon[12]{} ignites vigorously at the center in a radiative region 
(Fig. \ref{fig:kips} top-left) and burns outwards until it begins to drive a convective
\carbon[12]{} burning shell.
The star will then ignite oxygen in the 
core (non-explosively) and proceed through silicon and iron burning before 
collapsing in a core collapse. 

As the \crate{} rate increases the initial
\carbon[12]{} ignition point, defined where the nuclear energy generated is greater than the energy lost in neutrinos,
 moves outwards in mass coordinate 
(Fig. \ref{fig:kips} top-center).  As the \carbon{} abundance decreases, the star requires a higher density to burn
\carbon{} vigorously, thus the star must contract further which increases the neutrino losses.
No convective carbon shell forms before the oxygen in 
the core ignites explosively,
proceeding to a PPISN. Once the \crate{} rate increases sufficiently such that 
the core is depleted in \carbon[12]{} after core helium burning,
no \carbon[12]{} burning region forms and the core proceeds to 
ignite oxygen explosively (Fig. \ref{fig:kips} top-right) as a PISN leaving behind no black hole remnant.

Stars with a convective \carbon[12]{} burning shell can resist the 
collapse caused by the production of \pair{}, and thus maintain hydrostatic equilibrium until core collapse.
Figure \ref{fig:kips} (middle-left) shows that when the shell forms it prevents the center of the star
from reaching $\Gamma_{1}-4/3 < 0$. Therefore, the instability is only local and never becomes global: only a small 
region around  the carbon shell becomes unstable.
For stars without the convective carbon shell (middle-center and middle-right), a significant fraction of the entire
star becomes unstable resulting in a global instability.

Carbon burning begins at the center and moves outwards (either vigorously or not), thus
depleting the center of \carbon{}. Therefore, the carbon shell (if it forms) can not move inwards
as there is insufficient fuel for it to burn.
The region undergoing carbon burning  
can not move outwards either, as the convection zone is mixing the energy released
from the nuclear reactions over a significant portion of the star. This prevents layers above
the carbon burning region from reaching sufficient temperatures and densities needed for
vigorous carbon burning \citep{farmer15,takahashi18}.

The convective carbon shell can only be sustained then if
it can bring fresh fuel in via 
convective mixing from the rest of the core. 
Thus when a convective carbon shell forms it also allows additional 
fuel to be mixed into the burning region from the outer layers 
of the core. This prolongs the lifetime of the
carbon burning shell and prevents the collapse due to \pair{} from occurring, until the
carbon shell convective region is depleted in \carbon{} which may not occur before the
star undergoes core collapse.

As the carbon fraction decreases, the carbon shell burning becomes 
less energetic (due to the $\carbon[12]+\carbon[12]$ reaction
depending on the density of carbon squared).
Therefore, as $\csigma{}$ increases (and \carbon{} fraction decreases)
less energy is released from the carbon burning, thus  the fraction of the core where $\Gamma_{1}-4/3 < 0$ increases
There becomes a critical point where the carbon burning is insufficient  
to prevent the violent ignition of oxygen in the core, thus pulsations begin.
Around this critical region, convective carbon burning can still occur, however 
the burning region can undergo flashes, where the \carbon{} ignites but is then quenched.
This leads to weaker and shorter lived convection zones which do not mix in sufficient \carbon{}
to sustain a continuous carbon burning shell. This leads to very weak pulses removing only a few tenths
of a solar mass of material. For stars with these weak convection zones the carbon 
shell only delays the ignition of oxygen; once carbon 
is sufficiently depleted the oxygen can ignite explosively.
Eventually no carbon shell convection zone is formed at all (Fig. \ref{fig:kips} middle-center), this leads to larger
pulses removing solar masses of material.

The bottom row of Fig. \ref{fig:kips} shows the temperature-density profile inside the star at moments marked
in the top row of Fig. \ref{fig:kips}. As the stars evolve the core contracts and heats up, eventually the central regions of the star enter
the instability region ($\Gamma_{1}-4/3 < 0$). Once a convective carbon shell forms (bottom-left) the core
stops contracting homogeneously (along the line where the radiation pressure is equal to the gas pressure) and moves
to higher densities. This is due to the continued loss of entropy to neutrinos from the core. Thus
when oxygen ignites the core is now outside the instability region ($\Gamma_{1}-4/3 > 0$) and as such does not undergo
pulsational mass loss \citep{takahashi18}. 

For stars without a convective carbon shell (bottom-center and bottom-right)
the core continues to contract homogeneously. When oxygen ignites it does so inside the $\Gamma_{1}-4/3 < 0$ region.
As the temperature increases, due to the oxygen burning, the production of \pair{} increases causing a positive feedback loop.
This leads to the explosive ignition needed to drive a pulse. Stars undergoing a PPISN (bottom-center) have slightly lower
core entropies than stars undergoing a PISN (bottom-right) due to the small amount of non-convective carbon burning that
occurs before oxygen burning begins.

Further decreases in the carbon abundance leave 
little carbon fuel to burn. Thus as the star 
collapses due to the production of \pair{}, the oxygen
is free to ignite violently. This causes the star to undergo a PISN, 
completely disrupting the star. This can also be seen in
Figure \ref{fig:c12_posthe2d} where the boundaries between the different final fates
move to lower masses as \csigma{} increases, as the pulses become more
energetic for a given initial mass. 

\subsection{Black holes 
  above the PISN mass gap}\label{sec:abovegap}
  Figure \ref{fig:final_state} shows population of black holes that form above the PISN mass gap.
  These black holes form due to the failure of the PISN explosion to
  fully unbind the star \citep[e.g.,][]{bond84}.
  As the helium core mass is increased (at constant \csigma{}) a PISN explosion increases in energy, 
  due to a greater fraction of the
  oxygen being burnt in the oxygen ignition. This increased energy leads to an increase in the 
  maximum core temperature
  the star reaches, before the inward collapse is reverted and the star becomes unbound. 
  This increased temperature can be seen in the increased production of \nickel{} as the initial mass
  increases \citep{woosley:02, renzo20csm}. Eventually the core reaches sufficent temperatures that the energy extracted from the
  core by photo disintegrations is sufficent to prevent the star from becoming unbound.

  Figure \ref{fig:final_state} shows that as \csigma{} increases the initial helium core mass needed
  to form a black hole above the gap decreases. This is due to the increased production of oxygen
  as \csigma{} increases. At these masses we do not see the formation of a convective carbon shell,
  even for $\csigma=-3$, though some radiative carbon burning occurs (similar to point C in the middle panels of
  Figure \ref{fig:kips}).  Instead the cores with greater total amounts of oxygen can liberate
  greater amounts of energy from the oxygen burning. This burning raises the temperature in the core, allowing
  additional nuclosynethesis to occur with the silicon and iron group elements produced from the oxygen burning. 
  For a fixed initial helium core mass as \csigma{} increases, the peak core temperature increases.
  Once a core reaches 
  $\log\ \rm{T/K}\approx9.86$, then the rate of photodisintegrations is sufficient to prevent the star from
  unbinding itself. This is what sets the upper edge of the mass gap.

\section{Edges of the PISN mass gap}\label{sec:edgegap}

\begin{figure}[htp]
  \centering

  \includegraphics[width=\linewidth]{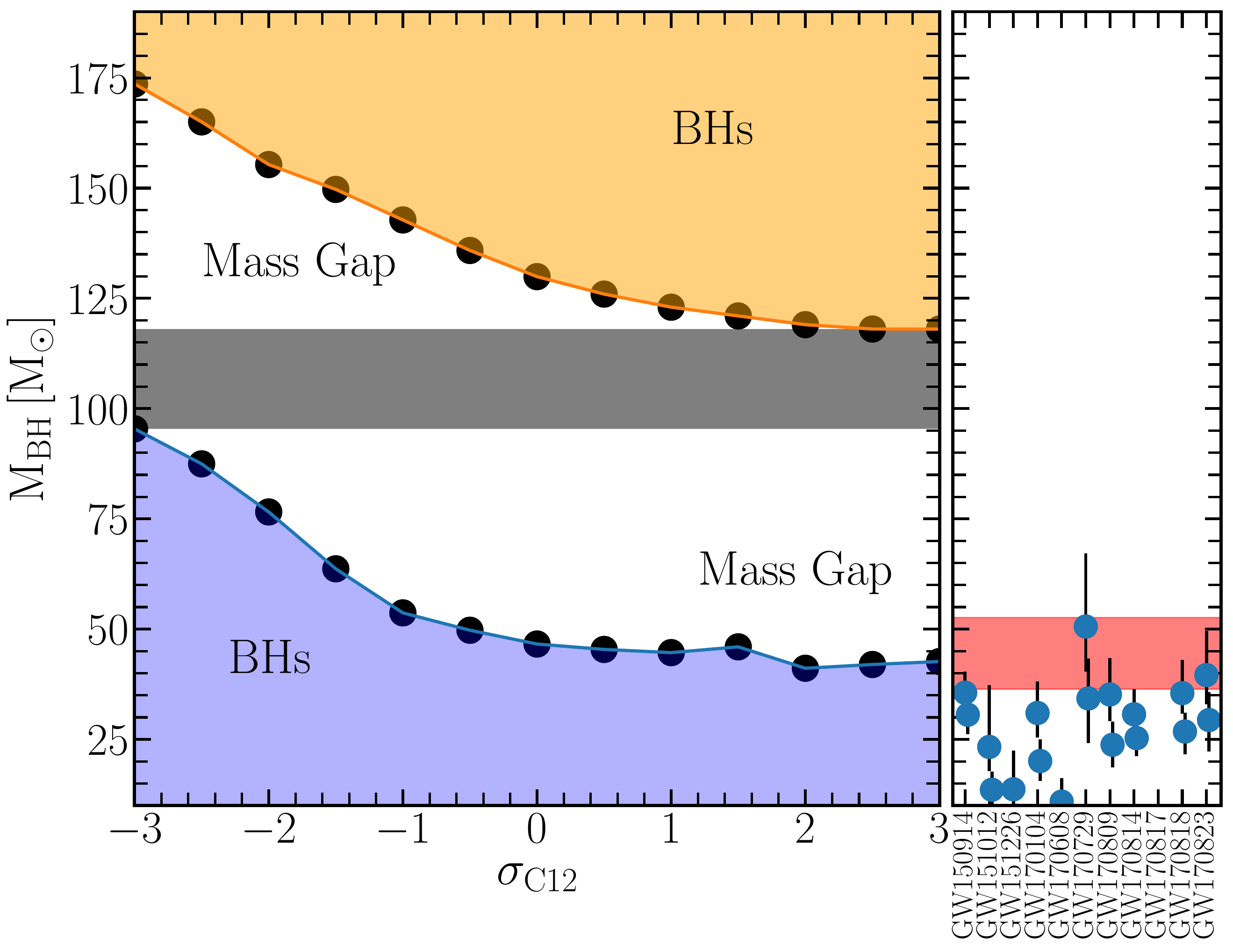}

  \caption{ The location of the PISN mass gap as a function of the 
  temperature-dependent uncertainty
  in the \crate{} reaction rate. The
  white region denotes the mass gap, while the grey horizontal bar region denotes the
   mass range where we can not place
  a black hole for any value of the \crate{} rate. The side plot shows the inferred masses of 
  the currently known black holes from LIGO/Virgo
  O2 \citep{ligo18b}, with their 90\% confidence intervals. The red region shows
  the 90\% confidence range for the inferred location of the lower edge of the PISN mass gap  from the O1/O2 data \citep{ligo18a}.}
    \label{fig:max_bh_c12sigma}
\end{figure}

Figure \ref{fig:max_bh_c12sigma} shows the location of the PISN black hole mass gap as a function of the
temperature-dependent uncertainty in \crate{}. As the rate increases 
(with increasing \csigma) both
the lower and upper edge of the PISN mass gap shift to lower masses, from 
$\approx90\msun$ to $\approx40\msun$ for the lower edge
and $\approx175\msun$ to $\approx120\msun$ for the upper edge.
The width of the region remains approximately constant at $83^{+5}_{-8}\msun$. 
The typical quoted value for the maximum mass of a black hole below the PISN mas gap is $45-55\msun$  \citep{yoshida:16,woosley:17,leung19,marchant:19,farmer:19}
The gray box in Fig. \ref{fig:max_bh_c12sigma}
shows the region of black hole masses, between $\mbh\approx90-120\msun$, where 
we can not place a black hole from a first generation core collapse or PPISN model.
Thus black holes detected in this mass region would need to come from alternate 
formation mechanisms, for instance; either second generation
mergers \citep{rodriguez16a,rodriguez19,gerosa19}, primordial black holes 
\citep{carr:16,alihaimoud17}, or accretion on to the black hole \citep{dicarlo19b,roupas19,vanson20}.

The detection of the upper edge of the PISN mass gap ($\mbh>120\msun$) would provide a strong constraint on the 
\crate{} rate.
This edge has smaller numerical uncertainties associated 
with it, as it is defined only by a combination of
fundamental physics (nuclear reaction rates and the equation of state of 
an ionized gas) and does not depend on the 
complexities of modeling the hydrodynamical pulses which define the 
lower edge of the PISN mass gap. Mergers in this mass range
are expected to be rare due to the difficulty in producing sufficiently massive stars 
in close binaries \citep{belczynski:16}, however they may be detectable by third
generation gravitational wave detectors \citep{mangiagli19}.

\subsection{Other sources of the \crate{} rate}

Table \ref{tab:bhmass_src} shows the maximum black hole mass as a function of 
different sources for the \crate{} reaction. The \citet{sallaska13} \STARLIB{} rate is based
on that of \citet{kunz02}, however \STARLIB{} assumes rate probability density is
lognormal thus its median value is $\sqrt{\rm{R_{Low}}\times\rm{R_{High}}}$ (eqaution 17 \citet{sallaska13}),
where $\rm{R_{Low}}$ and $\rm{R_{High}}$ are from \cite{kunz02}. The maximum black mass for the \citet{deboer17} rate is computed using the
``adopted", ``lower", and ``upper" rates from Table XXV of \citet{deboer17}.
The lower edge of the black hole mass gap over the different sources is between $47-51\msun$, with 
an uncertainty on the maximum black hole mass of $<5\%$.
The upper edge varies between $130-136\msun$, with a similar 
uncertainty on the maximum black hole mass of $<5\%$. 

The small variations seen in the edges of the PISN mass gap, are due to the fact that
the different sources of the
\crate{} have been slowly converging over time on a S-factor between $140-160\kevbarns$ (See figure \ref{fig:sfactor}).
See Figure 26 of \citet{deboer17} for a review of how the uncertainty in the
different energy levels has improved since the 1970's.

\begin{deluxetable}{ccc}

  \tablehead{\colhead{Source} & \colhead{Lower [$\msun$]} & \colhead{Upper [$\msun$]} }
  
  \tablecaption{Location of the edges of the PISN mass gap for different sources of the \crate{} 
  reaction rate. Uncertainties quoted are $1\sigma$  where applicable.\label{tab:bhmass_src} }
  
  \startdata
  $1.7 \times$ \citet{caughlan88} & 49 & 135 \\
  \citet{angulo99} (\NACRE{}) & 49 & 130 \\
  \citet{kunz02}\tablenotemark{a} & 50 & 134 \\
  \citet{cyburt10} (\REACLIB{}) & 50 & 136 \\
  \citet{sallaska13} (\STARLIB{})\tablenotemark{b} & $47^{+7}_{-2}$ & $130^{+13}_{-7}$ \\
  \citet{deboer17}\tablenotemark{c} & $51^{+0}_{-4}$ & $134^{+5}_{-5}$ 
  \enddata
  \tablenotetext{a}{Based on the ``adopted" fitting coefficents in Table 5 of \citet{kunz02} }
  \tablenotetext{b}{The \STARLIB{} median rate is based on  $\sqrt{\rm{R_{Low}}\times\rm{R_{High}}}$ (eqaution 17 \citet{sallaska13}), 
  where the rates $\rm{R_{Low}}$ and $\rm{R_{High}}$ come from Table 5 of \citet{kunz02}} 
  \tablenotetext{c}{Based on the ``adopted", ``lower", and ``upper" rates from Table XXV of \citet{deboer17}}

  \end{deluxetable}

\begin{deluxetable*}{cccccccc}

\tablecaption{The relative change in the location of the upper and lower edge of the PISN mass gap, when varying both the \crate{}
reaction rate and either the $3\alpha$, \carbon[12]+\carbon[12], or \oxygen[16]+\oxygen[16] reaction rate, with respect to our default choices.
\label{tab:bhmass_rate}}
\tablecolumns{8}
\tablehead{\colhead{Rate} & \colhead{Uncertainty} & \multicolumn{6}{c}{\crate}  \\
 & & \multicolumn{2}{c}{$\csigma=-3$} & \multicolumn{2}{c}{$\csigma=0$} & \multicolumn{2}{c}{$\csigma=+3$} \\
 & & Lower & Upper & Lower  & Upper  & Lower & Upper }

\startdata
$3\alpha $ & $+1\sigma$  & \phn $-1.0\%$ & \phn $-2.7\%$ & \phs \phn $1.7\%$\tablenotemark{a} & \phn $-0.8\%$ & \phn $-4.3\%$\tablenotemark{a} & \phn $-0.8\%$ \\
 \citet{sallaska13} & $-1\sigma$ & \phs \phn $6.2\%$ & \phs \phn $4.2\%$ & \phs \phn $8.9\%$\tablenotemark{a} & \phs \phn $4.6\%$ & \phn $-7.6\%$\tablenotemark{a} & \phs \phn $0.8\%$ \\
\hline
$\carbon[12]+\carbon[12]$ & $+1\sigma$  & \phn $-1.0\%$ & \phs \phn $0.5\%$\tablenotemark{a} & \phs $16.6\%$ & \phs $\sim0.0\%$ & \phn $-16.8\%$\tablenotemark{a} &  \phs $\sim0.0\%$ \\
\citet{tumino18}  & $-1\sigma$  & \phs \phn $1.9\%$ & \phs \phn $0.5\%$\tablenotemark{a} & \phn $-1.4\%$ & \phs $\sim0.0\%$ & \phn $-1.2\%$\tablenotemark{a} &  \phs $\sim0.0\%$ \\
\hline
$\oxygen[16]+\oxygen[16]$ &  $\times10$  & \phs \phn $0.0\%$\tablenotemark{b} & \phn $-1.8\%$ & \phn \phn $0.0\%$ & \phn $-2.3\%$ & \phn $-2.6\%$ & \phn $-3.3\%$ \\
 \citet{caughlan88}&  $\times0.1$ & \phs \phn $0.0\%$\tablenotemark{b} & \phs \phn $4.4\%$ & \phs \phn $5.7\%$ & \phs \phn $9.1\%$ & \phs \phn $6.9\%$ & \phs \phn $9.2\%$ \\
\enddata
\tablenotetext{a}{Variations have the same sign as numerical difficulties prevent comparison between similar models.}
\tablenotetext{b}{No variation seen as most burning occurs at temperatures where the rate has reverted to \citet{caughlan88} and thus shows no variation. }
\end{deluxetable*}

\subsection{Sensitivity to other reaction rates}

Table \ref{tab:bhmass_rate} shows how the maximum black hole mass varies as a 
function of both the \crate{} rate
and other reaction rates
that either create carbon (the $3\alpha$ reaction), destroy carbon ($\carbon[12]+\carbon[12]$), 
and oxygen burning ($\oxygen[16]+\oxygen[16]$). For each rate varied we compute the 
location of the mass gap for the number of standard deviations from the median for that rate, and 
for variations in the \crate{} reaction. This is to probe for correlations between the rates.
In the case of the rates from \citet{caughlan88} we follow the uncertainty provided by \STARLIB{} and we multiply (divide) the rate by a fixed factor of 10
due to the lack of knowledge of the uncertainty in this rate.
Table \ref{tab:bhmass_rate} then shows the fractional change in the location of each edge of the 
mass gap, as an indication of how sensitive the edges of the mass gap are to other uncertainties.
In general the maximum fractional error is $\approx5-15\%$ from considering other reactions, 
in the location of the PISN mass gap.

We would expect that varying a rate between
its upper and lower limits would produce relative changes with opposing signs, however in some cases this
does not occur. This is due to numerical difficulties in the evolution of the models. 
Some of these models would fail to reach core collapse, and thus we could not measure the peak of the black hole mass distribution. Instead,
  we report the largest black hole mass only for the models that successfully reach core collapse. 
  This means that the relative change we measure includes changes in the initial mass as well as the black hole mass.
Table \ref{tab:bhmass_rate} should therefore be taken as a representation of the changes expected for different rates, but
it does not show the complete picture. 

For the 
\STARLIB{} median \crate{} reaction rate,
the $3\alpha$ reaction produces a fractional uncertainty of $1-10\%$, independent
of the \crate{} rate. By increasing the $3\alpha$, within its one sigma uncertainties, rate we decrease the
maximum black hole mass. There is a much larger change when reducing the rate than when the rate
is increased. 
 
To test variations in the $\carbon[12]+\carbon[12]$ we use the rate provided by 
\citet{tumino18}, which provides a temperature dependent $1\sigma$ uncertainty on the reaction rate.
However, the uncertainty is only available up to $3\gk$, at higher temperatures we revert to
\MESA's standard $\carbon[12]+\carbon[12]$ reaction rate, which does not have a provided uncertainty estimate \citep{caughlan88}.

It is difficult to determine the trend in black hole mass compared to the $\carbon[12]+\carbon[12]$
given a number of models due not converge. We might expect variations around  $\approx 15\%$. 
The larger
change occurs when the $\csigma \geq 0$ than when $\csigma =-3$. This is due to the change in the power generated during the 
carbon burning. When the $\carbon[12]+\carbon[12]$ rate is increased then stars with low \carbon{} fractions ($\csigma \geq 0$),
which would not generate a convective carbon shell (when $\carbon[12]+\carbon[12]$ is small)
can now generate sufficient power to alter the core structure and potentially drive the formation of convective carbon burning shell.
See however \citet{tan20} for a discussion on why the \citet{tumino18} rate may have been overestimated, and
\citet{fruet20} for a discussion on new measurement techniques of the $\carbon[12]+\carbon[12]$ rate.

As the $\oxygen[16]+\oxygen[16]$ rate increases the maximum black hole mass decreases.
This change is asymmetric, with a larger change occurring when the rate decreases than when the rate increases.
The 0\% change seen when $\csigma=-3$, is due to those stars lacking pulsations. 
As the star has a high \carbon{}
fraction, and thus a carbon shell, it does not undergo explosive oxygen burning only stable core oxygen burning.
As there are no pulsations no mass is lost. This rate does have an effect on the core structure of the star,
which might lead to variations in the mass lost during the final collapse into a black hole.
For larger values of \csigma{} there is up to a $10\%$ variation in the location of the edges of the PISN mass gap.

More work is needed to understand the correlations
between the different reaction rates and their effect on our ability to constrain
the edges of the PISN mass gap, e.g. \citet{west13}. We need to improve our understanding of how the 
uncertainty in the rates at different temperatures 
alters the behavior of the carbon shell and the final black hole mass.
This could be achieved with a Monto-Carlo sampling
of the reactions rates, e.g. \citet{rauscher16,fields16,fields18}, however this comes at a 
much greater computational cost.

\begin{figure}[htp]
  \centering

  \includegraphics[width=\linewidth]{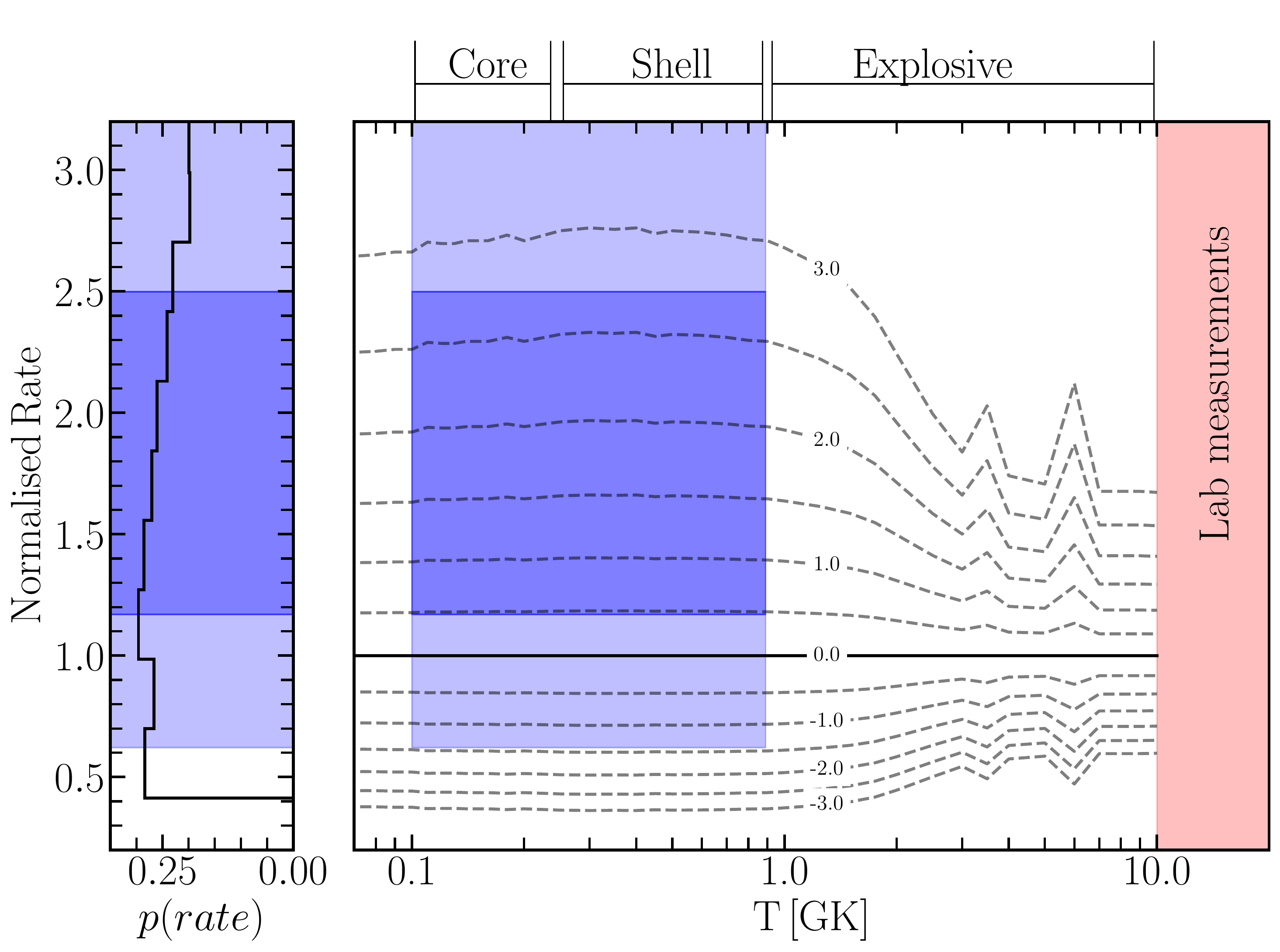}

  \caption{ 
  The right pannel shows 
  the \crate{} rate as a function of temperature, normalized to the median \STARLIB{} rate 
  $\langle \sigma \nu \rangle / \langle \sigma \nu \rangle_{\rm{median}} = e^{\mu\left(T\right)\csigma}$.
  The left panel shows the posterior of our distribution.
  The solid line, labeled 0.0, indicates the normalized median rate (i.e., \csigma=0.0). The dashed
  lines show the reaction rates above and below the median \STARLIB{} rate, labeled by the appropriate value
  of \csigma.
  We assume that the black hole mass distribution is given by model B of \citet{ligo18a}. The dark blue
  region shows the 50\% confidence range in the \crate{} rate, while the lighter blue shows the 90\% 
  confidence interval.
  Note the upper rate limit is unbounded when adopting the current O2 LIGO/Virgo
   posteriors on the maximum black hole mass.
   The red region ($\rm{T}>10\,GK$) shows the approximate lower edge of the energy range for 
   lab measurements of the \crate{} rate \citep{holt19}.}
    \label{fig:rate_limits}
\end{figure}

\section{Constraining the \crate{} reaction rate with gravitational waves}\label{sec:rates}

Because of the sensitivity of the edges of the PISN mass gap to the \crate{} reaction rate, we can use
the measured location of the gap to derive a value for the \crate{} at
astrophysically relevant temperatures. See Appendix 
\ref{sec:c12_depend} for the sensitivity of our results to different temperature ranges.
We focus here on the lower edge of the mass gap, as it has been inferred from the existing 
LIGO/Virgo data \citep{ligo18a}.

The currently most massive black hole in O1/O2, as inferred by LIGO/Virgo is GW170729 at 
$\mbh=50.6^{+16.6}_{-10.2}\,\msun$ \citep{ligo18a}, which could be used as an estimate
for the location of the PISN mass gap,
assuming that it is from a first generation black hole. There are also several other 
candidates for the most massive black hole.
This includes 
IMBHC-170502 has which has been inferred to have individual black hole components with 
masses $\approx94\msun$ and $\approx62\msun$ \citep{udall19}. 
GW151205 has also been proposed 
to have one component with an inferred mass of $\mbh=68^{+28}_{-17}\msun$ \citep{Nitz20}. 
By having a component mass inside the classical PISN mass gap it was suggested that this 
was the result of dynamical mergers. 

However, we must be careful in not over-interpreting single events,
which may be susceptible to noise fluctuations \citep{fishbach20} which can
make a black hole have a higher apparent mass than it truly does.  
For instance, considering GW170729 jointly with the other O1/O2 detections, lowers
its mass to $\mbh=38.9^{+7.3}_{-4.5}\msun$, which places it below the PISN mass gap.

Thus we must consider the entire population of binary black hole mergers as a whole,
when measuring the maximum inferred black hole mass below the gap. The 10 detections in O1/O2
places the maximum black hole mass below the PISN mass gap at $42-44\msun$ depending
on the choice of model parameters \citep{ligo18a}. The current 90\%
confidence interval on this value is $\approx\pm10\msun$.
With a large enough population ($\mathcal{O}(50)$) of black holes we can place limits of
$\approx \pm1\msun$ on the location of the gap \citep{fishbach:17,ligo18a}.

We assume that all binary black hole mergers so far detected come from isolated binaries
or first generation black hole mergers, thus the maximum mass black holes below the gap 
come from PPISN. We also
assume that only uncertainties 
in \crate{} matter. Thus we can use
the posterior distribution over the maximum black hole mass for the population of black holes as the 
estimate of the maximum black hole mass below the mass gap.

Figure \ref{fig:rate_limits} shows the uncertainty in the \STARLIB{} 
\crate{} reaction rate as a function of temperature.
Over the temperatures we are sensitive to, less than $1.0\gk$ where helium burns non-explosively, 
the uncertainty is approximately constant.
Thus we need only find a single temperature-independent $\csigma$ to 
fit to the maximum black hole mass below the gap.

We fit a 4\ts{th} order polynomial to the lower edge of the PISN 
mass gap (Fig. \ref{fig:max_bh_c12sigma}) to
map from maximum black hole mass to $\csigma$.
This is then combined with the posterior of the maximum black hole mass
to generate a posterior distribution
over $\csigma$. The blue boxes in Fig. \ref{fig:rate_limits} 
show our 50\% and 90\% confidence interval
on $\csigma$. At 50\% confidence, we can limit $\csigma$ 
to be between 0.5 and 2.5$\csigma$, while
at 90\% confidence we can only place a lower limit of $\csigma>-1.5$. 
This is due to the posterior distribution from
LIGO/Virgo allowing the maximum black hole mass to be $\maxbh<40\msun$, which is 
below the lower limit we find for the edge of the mass gap. 

By taking the \STARLIB{} astrophysical S-factor, at $300\kev$, to be 
$165\kevbarns$ \citep{sallaska13} we can scale the S-factor from the $\csigma$.
As the normalized rate is approximately flat (Fig. \ref{fig:rate_limits}) the 
uncertainty is flat. 
Using Equation \ref{eq:sfactor} and Fig. \ref{fig:rate_limits} we have
$e^{\mu\left(T\right)\csigma} = \langle \sigma \nu \rangle / \langle \sigma \nu \rangle_{\rm{median}} \propto S(300\,\kev) $
,
thus the S-factor can be linearly scaled from its 
\STARLIB{} value to a new value, given by a different \csigma{}.

\begin{figure}[htp]
  \centering

  \includegraphics[width=\linewidth]{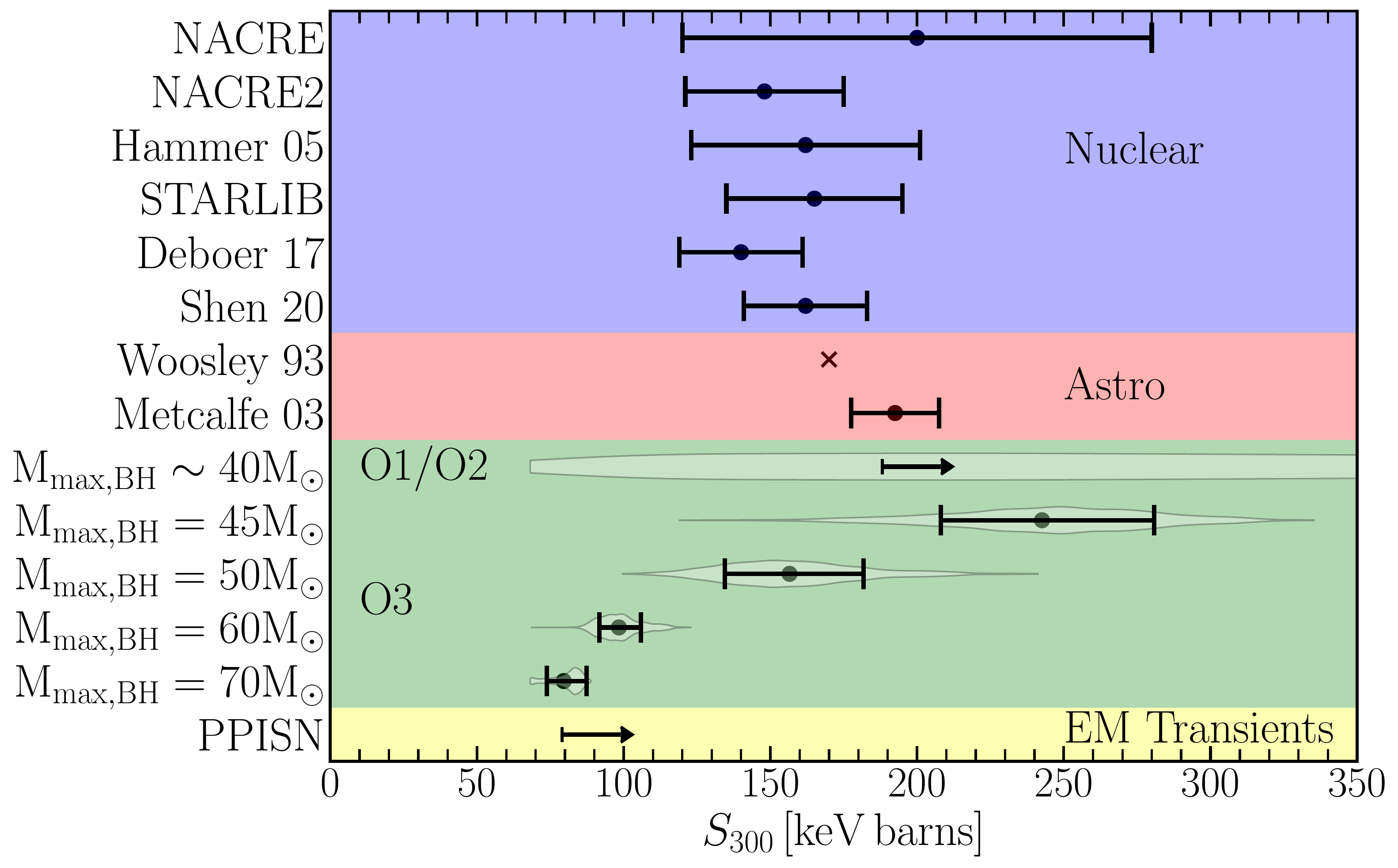}

  \caption{Constraints on the S-factor at $300\kev$ for the \crate{} reaction rate.
   The blue region presents values for commonly-used nuclear reaction rate libraries,
    while the red region shows values
  inferred from astrophysical measurements, by observations of white dwarfs \citep{metcalfe03}
  and from the galactic chemical enrichment model of \citet{woosley93}. 
    The green region show our posteriors on the \crate{} derived from the O1/O2
    maximum black hole mass inferred from model B of \citet{ligo18a}, and 
    for 50 simulated binary black-hole observations, approximately the number expected in O3, 
    assuming that the O3 detections follow the power law model B from \citet{ligo18a}.
    Error bars show the 68\% confidence intervals on our inferred S-factor at $300\kev$. 
  The yellow region shows the constraints placed if PPISN supernovae exist (Section \ref{sec:sn}). 
  Error bars in the blue and red regions are $1\sigma$ uncertainties.
  We assume that the uncertainty from \citet{shen20} is the same as that of \citet{deboer17}.
  }
    \label{fig:sfactor}
\end{figure}

Figure \ref{fig:sfactor} shows a 
comparison between the values
for the S-factor from nuclear laboratory experiments, constraints 
placed by white dwarf asterosiesmology and galactic chemical
enrichment models, and this study. For our assumptions we find
the S-factor for the \crate{} rate at $300\kev$ to be 
$S_{300}>175\kevbarns$, at 68\% confidence.
At 95\% confidence we find a limit of  $S_{300}>82\kevbarns$ and at 
 99\% confidence we find a limit of  $S_{300}>68\kevbarns$.
The S-factors computed here are consistent with experimentally derived 
values, though we only currently place a lower limit on the S-factor.
See  section \ref{sec:future} for a discussion on how this limit may be improved with future
gravitational wave detections. 
Figure \ref{fig:sfactor} also shows the lower limit on the S-factor if
PPISN can be shown to exist though 
other means, for instance electromagnetic observations of the SN. The 
existence of PPISN would imply that
$S_{300}>79\kevbarns$, this is discussed further in Section \ref{sec:sn}.

\subsection{GW190521}

The detection of GW190521 \citep{GW190521a} with component masses of 
$85^{+21}_{-14}\msun$ and $66^{+17}_{-18}\msun$ places \emph{both} black holes firmly in the 
mass gap, as inferred from the O1/O2 observations \citep{GW190521b}. However, if we assume that the 
primary black hole in GW190521 was a first generation black hole, this leads to an inference of 
$\csigma=-2.4^{+0.6}$, and thus $S(300\,\kev)=73^{+11}\kevbarns$. We can not infer a lower limit 
as the 90\% confidence interval for the mass extends above $\mbh\approx95\msun$ where our models 
can not place a black hole, below the mass gap. If $\csigma\approx-2.5$, then 
PPISN would be suppressed due to the carbon burning shell. Instead, if we assume that only the 
secondary object was a first generation black hole we would infer a $\csigma=-1.7^{+1.8}_{-0.5}$, 
and thus $S(300\,\kev)=87^{+84}_{-12}\kevbarns$. 

However, it is unlikely that GW190521 is a pair of first generation black holes, instead it is 
likely to be a pair of second generation black holes, perhaps inside an AGN disk \citep{GW190521b}. 
Indeed, \citep{graham20} claims a tentative detection of EM counterpart (ZTF19abanrhr) due to the 
merger product ramming into an AGN disk. However, the redshift of the AGN does not agree with the 
GW-inferred redshift. Further investigation of this event and the likelihood of a double second 
generation merger in an AGN disk is warranted. The presence of black holes in the expected mass 
gap region can make the analysis of the location of the gap more difficult, though folding in a 
prior based on the \crate{} may help to identify outliers that have formed through alternative 
formation mechanisms.

\section{Prospects for constraining \crate{} from future GW detections}\label{sec:future}

With the release of the O3 data from LIGO/Virgo, it is predicted that there 
will be $\mathcal{O}(50)$ 
BBH detections \citep{abbott18}. Thus we can ask how well can we expect to do with 
additional observations? This 
depends strongly on what value the lower edge of the PISN mass gap is 
found to be.  

We assume that the detections in O3 will follow a power law in primary mass 
($\rm{M}_{1}$) of the black hole in the binary, with the form previously
assumed for the O1/O2 data \citep{ligo18a}. Thus we have:

\begin{equation}
p(\rm{M}_{1}) \propto \rm{M}_{1}^\alpha
\end{equation}

\noindent for $\rm{M_{min}} < \rm{M}_{1} < \rm{M_{max}}$, where $\rm{M_{min}}$ is the minimum possible black hole mass
below the PISN mass gap, 
and $\rm{M_{max}}$ is the maximum possible black hole mass.
We also assume that the secondary mass $\rm{M_{2}}$ follows:

\begin{equation}
p(\rm{M}_2 | \rm{M}_1) \propto \rm{M}_2^\beta
\end{equation}

\noindent where $\rm{M_{min}} < \rm{M}_2 < \rm{M}_1$.
This is equivalent to model B of \citet{ligo18a}. 
We set $\rm{M_{min}}=5\msun$
and $\beta=4$ to be consistent with the current O1/O2 observations  \citep{ligo18a}.
There is a 
strong correlation between
the values of $\alpha$ and $\rm{M_{max}}$. Thus for different choices of $\rm{M_{max}}$ 
we choose a
value of $\alpha$ that remains consistent
with the O1/O2 data (e.g, Fig. 3 of \citet{ligo18a}).
As $\rm{M_{max}}$ increases, we require a larger $\alpha$ value.
We consider 4 possible locations for the lower edge of the PISN mass gap ($\rm{M_{max}}$) to explore how the
uncertainty both from the LIGO/Virgo measurements and from our stellar models
affects our determination of the \crate{} rate.
We considered $\rm{M_{max}}=45, 50, 60, 70\msun$,
  and we choose $\alpha=-1.5, -1.5, -2, -3$, to remain consistent with the joint $\alpha, \rm{M_{max}}$ posterior 
  from O1/O2 \citep{ligo18a}.

We generate 50 mock detections 
from each of the four mass distributions described above. 
  We assume that the underlying merger rate density is constant in redshift, and that 
  sources are detected if they pass an SNR threshold of 8 in a single detector. 
  We neglect the spins of the black holes in this study, as the maximum mass is 
  well-measured independently of the underlying spin distribution.
  For each mock detection, we simulate the measurement uncertainty on the source-frame 
  masses according to the prescription in \citet{fishbach20}, which is 
  calibrated to the simulation study of \citet{vitale17}. 
  This gives a typical $1\sigma$ measurement uncertainty of 
  $\sim 30\%$ on the source-frame masses. We then perform a hierarchical Bayesian 
  analysis \citep{mandel10,mandel19} 
  on each set of 50 detections to recover the posterior over the population parameters, 
  $\alpha$, $M_\mathrm{max}$, $\beta$ and $M_\mathrm{min}$. 
  This provides a projection of 
  how well $M_\mathrm{max}$, the maximum mass below the PISN mass gap, can be measured with 
  50 observations. 
  We sample from the hierarchical Bayesian likelihood using \texttt{PyMC3} \citep{salvatier16}.
  Finally, we translate the projected measurement of 
  $M_\mathrm{max}$ under each of the simulated populations to a measurement of 
  the \crate{} reaction rate according to Fig. \ref{fig:max_bh_c12sigma}.

Figure \ref{fig:sfactor} shows our inferred S-factors (at $300\kev$) for different choices of the
maximum black hole mass below the PISN mass gap. As the maximum black hole mass increases, the S-factor 
decreases (as stars have more \carbon[12]{} in their core and thus have reduced mass loss from pulsations).
The 68\% confidence interval also reduces in size as the maximum black hole mass increases.
The predicted accuracy with which LIGO/Virgo is expected to infer the maximum black hole mass  
decreases as the mass increases, as we require a steeper power law index $\alpha$ to be consistent with the
O1/O2 observations. This leads to 
fewer mergers near the gap. However, the 
maximum black hole mass becomes more sensitive to the choice of $\csigma$. This 
can be seen 
in the gradient of the lower edge of the PISN mass gap in Fig. \ref{fig:max_bh_c12sigma}, which
increases as \csigma{} decreases.
We caution that we have likely under-estimated the size of the uncertainty range, especially at the higher 
black hole masses, due to the
effect of uncertainties in other reaction rates and mass lost during the formation of the black hole.
With the predicted accuracy expected for LIGO/Virgo during O3 in inferring the maximum black hole mass,
we will be limited by the accuracy of our models, not the data, in constraining the \crate{} reaction rate.

\section{Other observables}\label{sec:bhimf}

\subsection{Formation rates}

Given that the initial 
mass function (IMF) strongly favors
less massive progenitors, this would imply
that PPISN and PISN would be more common at higher values of \csigma{} (higher \crate{} rates), all else being equal. 
This is potentially detectable given a sufficiently large population
of binary black hole mergers or PPISN/PISN transients, and could provide additional constraints on the \crate{} rate.
A number of upcoming surveys, including \code{LSST}, \code{JWST}, and \code{WFIRST}, are expected to find 
significant numbers of PPISN and PISN transients \citep{young08,hummel12,whalen13a,wahlen13b,villar18,regos20}.

To provide a rough estimate for this, we make the simplified assumption that the helium core masses follow a Salpeter-like IMF with a power
law $\alpha=-2.35$, so that we can compare the relative difference in formation rates for stars
with $\csigma=\pm1$. 
We take the smallest helium cores to be $\mheint=30\msun$ (the least massive stars 
modeled in our grid), and the maximum 
helium core mass that makes a black hole as: $73\msun$ for $\csigma=-1$, and $61\msun$ for $\csigma=+1$.
This folds together the formation rates of black holes from both core collapse and PPISN, as
gravitional waves can not distinguish these objects from their gravitional wave signal alone.
This leads to a relative increase in the formation rates of black holes of
$\approx10\%$ for the $\csigma=-1$ over $\csigma=+1$ models.
There is a larger variation possible, if we consider how the lowest-mass helium core 
that makes a black hole varies with \crate{}. For $\csigma=\pm1$ this puts the lower limit
at $5\msun$ and $8\msun$ \citep{sukhbold20}, which would lead to a factor 2 difference in the 
formation rates of black holes. This is mostly due to the change in the relative number of low mass
black holes.

\begin{figure}[htp]
  \centering

  \includegraphics[width=\linewidth]{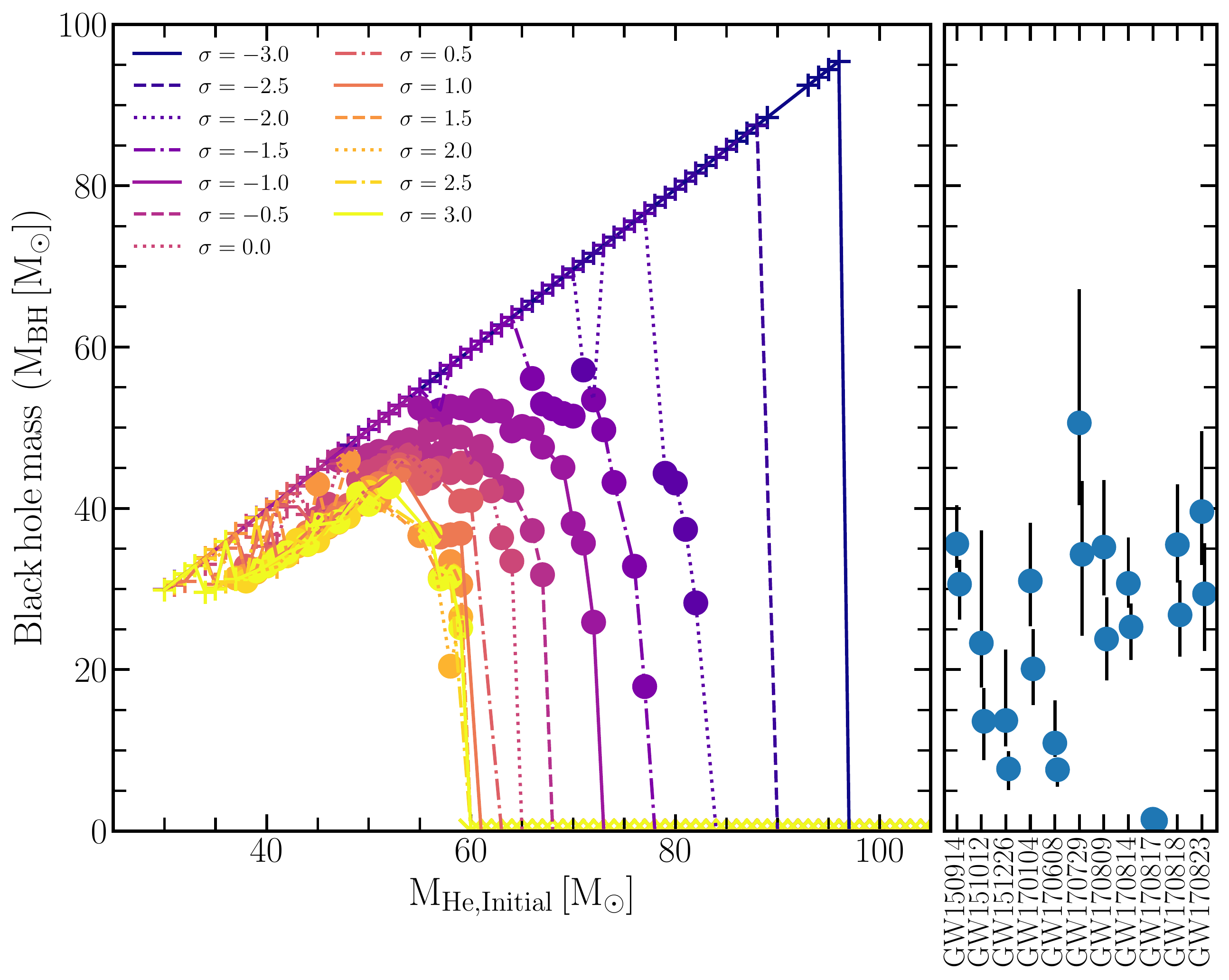}

  \caption{The final black hole mass as a function of the initial helium core mass. The colors
  represent different choices for \csigma{}. Circles denote models which undergo pulsational mass loss (PPISN)
  while plus symbols denote models which do not undergo mass lost due to pulsations (CC). The right pannel shows 
  the inferred masses of 
  the currently known black holes from LIGO/Virgo
  O2 \citep{ligo18b}, with their 90\% confidence intervals. }
    \label{fig:hedist}
\end{figure}

Figure \ref{fig:hedist} shows the black hole masses as a function of the initial helium core mass,
for different choices of \csigma{}. At the lowest values of \csigma{}, no star undergoes PPISN, 
thus the black hole mass scales lineraly with the initial helium core mass.
As \csigma{} increases the final black hole mass shifts to lower masses away from a linear relationship with the
helium core mass, due to pulsational mass loss. There is also a turn over in the initial-black hole mass realtionship,
where the most massive black holes do not form from the most massive stars undergoing PPISN \citep{farmer:19}.
This turn over may be detectable in the infered black hole mass distribution (see model C of \citealt{ligo18a}).
There may also be a small bump in the black hole mass distribution at the interface between the stars undergoing CC
and the lightest PPISN progenitors, depending on the strength of the mass loss in the lighest PPISN progenitors 
\citep{renzo20conv}. We chose not to show Figure \ref{fig:hedist} in terms of the carbon core mass, which is the more
applicable quantity to show when comparing between different
metalicities \citep{farmer:19}, as the highest \csigma{} models
have $X\left(^{12}\rm{C}\right)\approx0$ in their cores. This leads to the carbon core mass being ill-defined.

We can compare the formation rates of PPISN (and thus potentially detectable supernovae) only for $\csigma>-2.0$. 
As an indication of the variation, we consider the relative change (due to the IMF)
in formation rates only for $\csigma{}=\pm1$.
We find the helium core masses for stars that we would classify as undergoing a PPISN, as
$41-61\msun$ for $\csigma=+1$, and $55-73\msun$ for $\csigma=-1$. This leads to
a increase by a factor of 2 for $\csigma=+1$ over $\csigma=-1$. 
However,
there can be difficulty in determining which stars are PPISN
due to the small amounts of mass lost for the lightest PPISN progenitors \citep{renzo20csm}.

For the rate of PISN, we find the initial helium core masses to be $61-124\msun$ for $+1\csigma$, 
and $73-142\msun$ for $\csigma=-1$,
These ranges then lead to a relative increase of $\approx30\%$ for the rate of PISN for 
$\csigma=+1$ compared to $\csigma=-1$.
These variations are driven by the change in the lower masses needed for a PPISN or PISN as
the \crate{} rate increases \citep{takahashi18}.

\subsection{Observations of supernovae}\label{sec:sn}

In Fig. \ref{fig:final_state} we show that when $\csigma<-2$ no PPISN are formed.
This provides an intriguing
observational test for the \crate{} reaction rate. If the existence of PPISN
can be confirmed though photometric observations  (for instance candidate 
PPISN SN2006gy \citep{woosley:07}, SN2006jc \citep{pastorello07},
SN iPTF14hls \citep{arcavi:17, woosley:18,vigna19}, SN iPTF16eh \citep{lunnan:18}, or SN2016iet \citep{gomez19},
or SN2016aps \citep{nicholl20}) 
then we can place a lower limit
on the \crate{} reaction rate of $\csigma>-2.0$.

The outer layers of the star that are expelled in a pulsational mass loss event,
or even the small amount of material that might be ejected in the final collapse to
a black hole, provides some information on the final composition of the star,
 though this will depend on the star's metallicity and assumed wind mass loss rate.
In general the lowest mass stars that undergo a PPISN, in our grid, have surface layers dominated by
\helium{}, while the higher mass stars have \oxygen{} rich outer layers \citep{renzo20csm}.

As the initial mass increases, and the pulses get stronger and thus remove more mass
they can expose \oxygen{} rich layers, with traces of \neon{}, \sodium{}, \magnesium{}, and \silicon{} \citep{renzo20csm}.
However, the \carbon{} fraction in the outer layers follows the same trend seen in the core \carbon{}
fraction. As the \crate{} reaction rate increases more \carbon{} is converted into \oxygen{} in helium shell
burning. Therefore, the measured abundances could provide additional constraints on \crate{},
and may provide constraints in a reduced temperature region (that associated with shell helium burning,
$0.3<\rm{T}/\gk<1.0$). Further work is needed to quantify the amount of mixing in the outer layers of the star \citep{renzo20conv},
understanding which parts of the ejecta (and thus which layers of the star) would be measured in  
spectroscopy of a PPISN \citep{renzo20csm}, as well as investigating the effect of a larger nuclear network to follow
in greater detail the nucleosynthetic yields from the explosive oxygen burning \citep{weaver93b,heger02,woosley07,west13}.

\section{Discussion}\label{sec:discuss}

This work assumes that all black holes found by LIGO/Virgo so far have come from
stars that lose their hydrogen envelope before their collapse to a black hole. 
Thus the maximum mass a black hole can have is
limited by PISN and mass loss from PPISN.
However there are formation mechanisms which may place a black hole in the mass gap.
If a star can retain its hydrogen envelope until collapse,
though a combination of weak stellar winds \citep{woosley:17} or for a stellar merger 
\citep{vigna19,spera19}, then 
we could find black holes up to $\sim60\msun$, see \citet{vanson20} for an overview. 
Black holes formed in dense stellar clusters,
or AGN disks \citep{McKernan18}, can merge multiple times \citep{rodriguez16a, stone:17, dicarlo19,yang19}.

The first generation black holes in these environments will be limited by the PISN mass gap,
however higher generation mergers would not be limited and could populate the gap \citep{dicarlo20}.
Their effect on the inference of \crate{} will depend on
 whether the kick the resulting black hole receives is small
enough that the black hole stays in the cluster \citep{Rodriguez18,fragione18},
and thus on their uncertain contribution to the total
rate of black hole mergers. If they are distinguishable from mergers due to 
isolated binary evolution, for instance via their spin \citep{fishbach17a,gerosa17,Bouffanais19,sedda20}, then they could be removed from the population used to infer the \crate{} rate.
It may also be possible to fit the maximum black hole mass below the gap assuming the population contains both isolated binaries and hierarchical mergers without needing to subtract out the hierarchical mergers \citep{kimball20} 
Also, if a channel that produces black holes in the mass gap is rare, we may still be able to determine the location of the mass gap for the more dominant channel.

In this work we have considered four reaction rates, $3\alpha$, \crate{}, $\carbon+\carbon$, and 
$\oxygen+\oxygen$. There are many other reaction rates that can alter the evolution of a
star  \citep{rauscher16,fields18}. We expect their effect to be small for the final black hole mass,
compared with the reaction rates considered here, though they would play a role in the nuclosynethic yields 
from the PPISN and PISN ejecta.
In \citet{farmer:19} we investigated other uncertainties, e.g mass loss, metallicity,
convective mixing, and neutrino physics, have only a small effect of $\approx10\%$, on the location
of the mass gap. How convection is treated 
in this hydrodynamical regime
can have a small impact on the final black hole mass for stars at the boundary between
core collapse and PPISN i.e, where the pulses are weak. However the edge of the PISN mass gap
is not affected, due to the stronger pulses experienced by a star near the mass gap \citep{renzo20conv}.
In this work we considered only
non-rotating models. Rotation may play a role in the final
black hole mass, depending on how material with high angular momentum is accreted onto
the black hole during the collapse \citep{fryer04,rockefeller:06,batta17}.

By assuming the entire bound mass of the star collapses into a black hole, we place an upper limit on the black
hole mass possible for a PPISN. If the star was to lose mass during the collapse, 
then we would over estimate the inferred S-factor for the \crate{} reaction rate. 
If the star forms a proto-neutron star during its collapse it might eject $\approx10\%$ of the proto-neutron star mass
as neutrinos, $\approx0.1\msun$ \citep{fryer99,fryer:12}. The envelope of the star may then respond to the change in the gravitational potential,
generating a weak shock that unbinds material with binding energies less than $10^{47}\ergs$ \citep{nadezhin:80,lovegrove:13,fernandez18}. 
However, stars undergoing PPISN
will have already expelled their weakly bound outer envelopes, thus mass lost via a weak shock is limited to a few tenths of a solar mass.
If a jet is produced, by accretion onto the compact object, then there maybe an ejection $>1\msun$ of material \citep{gilkis16,quataert19}.
Assuming $1\msun$ of mass loss during the collapse, then our inferred S-factor at 68\% confidence decreases from 
$S_{300}>175\kevbarns$ to $S_{300}>159\kevbarns$. Further work is needed to understand the collapse
mechanism of these massive cores, and whether we can extrapolate from models of stars that core collapse with $5-10\msun$
cores to those with $\approx50\msun$ cores.

Previous studies of PPISN and PISN progenitors have found the location of the PISN mass gap consistent with ours,
$50-53\msun$ \citep{yoshida:16}, $\approx48\msun$ \citep{woosley:17}, $\approx50\msun$ \citep{leung19}.
The small variations in the location of the gap can be attributed to differences 
in chosen metallicity, mass loss rates, and source of the \crate{} reaction rate.
\citet{takahashi18} showed
how increasing the reaction rate decreases the initial mass needed for PISN,
in agreement with our findings and that the range of initial masses that can
form a PPISN is reduced as the reaction rate decreases.

\section{Summary}\label{sec:conclusion}

With the rapid increase in the number of gravitational-wave detections, the hope is that they can be used to
start drawing lessons about the uncertain physics of their massive-star progenitors.  
In an earlier paper \citep{farmer:19} we speculated that measurements of the edge of the predicted black-hole mass gap due to pair-instability could be used to constraint the nuclear reaction rate of carbon capturing alpha particles producing oxygen (\crate{}). This reaction rate is very uncertain but has large astrophysical significance. It is crucial in determining the final properties and fate of a star, and as we explicitly show in this work the predictions for the location of the pair-instability mass gap.  

We show that the physical reason why this reaction rate is so important is that it determines the relative fraction of carbon and oxygen in the core at the end of helium burning. In models for which we adopted a lower reaction rate, enough carbon is left to ignite such that it effectively delays the ignition of oxygen. As carbon-carbon burning occurs in a shell, the core inside the shell contracts to higher densities. This increases the effects of electron degeneracy and gas pressure, which stabilizes the core. The formation of electron-positron pairs is then suppressed due to the increased occupation fraction of the low energy states for electrons. Oxygen can then ignite stably even for higher core masses.  
 
In contrast, in models for which we assume higher reaction rates, almost all carbon is depleted at the end of helium burning. The star then skips carbon-carbon burning and oxygen ignites explosively. The net effect is that increasing the \crate{} reaction rate pushes the mass regime for pair pulsations and pair-instability supernovae to lower masses. This allows for lower mass black holes and thus shifts the location of the pair instability mass gap to lower masses.

Our results can be summarized as follows:

\begin{enumerate}

\item {\bf The location of the gap is sensitive to the reaction rate for alpha captures on to carbon, but the width of the mass gap is not.}  The lower edge of the mass gap varies between $47^{+49}_{-4}\msun$ and $130^{+44}_{-12}\msun$ for the upper edge, for $\pm3$ sigma variations in the \crate{} reaction rate.  The width is $83^{+5}_{-8}\msun$  (Figure~\ref{fig:max_bh_c12sigma}).

\item {\bf We can place a lower limit on the \crate{} reaction rate using the first ten gravitational-wave detections of black holes.}  Considering only variations in this reaction, we constrain the astrophysical S-factor, which is a measure of the strength of the reaction rate, to $S_{300}>175\kevbarns${} at 68\% confidence 
   (Figure~\ref{fig:sfactor}).

\item {\bf  With $\mathcal{O}(50)$ detections, as expected after the completion of the third observing run, we expect to place constraints of $\pm$10--30\kevbarns{} on the \crate{} S-factor.}
We show how the constraints depend on the actual location of the gap (Figure~\ref{fig:sfactor}).   

\item {\bf We find other stellar model uncertainties to be subdominant, although this needs to be explored further.}  Variations in other nuclear reactions such as helium burning ($3\alpha$), carbon burning ($\carbon+\carbon$), and oxygen burning ($\oxygen+\oxygen$) contribute uncertainties of the order of 10\% to the edge of the mass gap  (Table~\ref{tab:bhmass_rate}). See \citet{farmer:19} and \citet{renzo20conv} for a discussion of the effect of physical uncertainties and numerics.

\item {\bf The unambiguous detection of pulsational pair-instability supernovae in electromagnetic transient surveys would place an independent constraint on the \crate{} reaction rate.} For the lowest adopted reaction rates ($< -2\sigma$) we no longer see pulsations due to pair instability.  The detection of pulsational pair-instability would thus imply  a lower limit of $S_{300}>79\kevbarns$ for the \crate{} reaction rate (Figure~\ref{fig:final_state}, Section~\ref{sec:sn}). This will be of interest for automated wide-field transient searches such as the Legacy Survey of Space and Time (LSST). 

\item {\bf Constraining nuclear stellar astrophysics is an interesting science case for third generation gravitational wave detectors.} Future detectors such as the Einstein Telescope and Cosmic Explorer will be able to probe detailed features in the black-hole mass distribution as a function of redshift, and potentially lead to detections above the mass gap.  Improved progenitor models will be needed to maximize the science return as the observational constraints improve,  but the future is promising.   

\end{enumerate}

\acknowledgments
We acknowledge helpful discussions with B.~Paxton, F.~Timmes, P.~Marchant, 
E.~C.~Laplace, and L.~van Son.
RF is supported by the Netherlands Organization for Scientific Research (NWO)
through a top module 2 grant with project number 614.001.501 (PI de Mink).
SdM acknowledges funding by the European Union's Horizon 2020 research
and innovation program from the European Research Council (ERC)
(Grant agreement No.\ 715063), and by the Netherlands Organization for
Scientific Research (NWO) as part of the Vidi research program BinWaves
with project number 639.042.728.
SdM also acknowledges the Black Hole Initiative at Harvard University, 
which is funded by grants from the John Templeton Foundation and the Gordon and Betty Moore 
Foundation to Harvard University.
This work was also supported by the Cost Action Program ChETEC CA16117.
This work was carried out on the Dutch national e-infrastructure
with the support of SURF Cooperative. This research has made use of NASA's
Astrophysics Data System.

\appendix
\section{\MESA{}}\label{sec:appen_mesa}

When solving the stellar structure equations \MESA{} uses a 
set of input microphysics.
This includes 
thermal neutrino energy losses from the fitting formula of \citet{itoh96}.
The equation of state (EOS) which is a blend of the OPAL \citep{Rogers2002}, SCVH
\citep{Saumon1995}, PTEH \citep{Pols1995}, HELM
\citep{Timmes2000}, and PC \citep{Potekhin2010} EOSes.
Radiative opacities are primarily drawn from OPAL \citep{Iglesias1993,
Iglesias1996}, with low-temperature data from \citet{Ferguson2005}
and the high-temperature, Compton-scattering dominated regime from
\citet{Buchler1976}.  Electron conduction opacities are taken from
\citet{Cassisi2007}.

\MESA's default nuclear reaction rates come from a combination
of \NACRE{} \citep{angulo99} and \REACLIB{} \citep{cyburt10} (\textit{default} snapshot%
\footnote{Dated 2017-10-20. Available from http://reaclib.jinaweb.org/library.php?action=viewsnapshots }).
The \MESA\ nuclear screening corrections are provided by \citet{chugunov2007}.
Weak reaction rates are based on the following
tabulations; \citet{langanke2000},
\citet{oda1994}, and
\citet{fuller1985}.

Reverse nuclear reaction rates are computed from detailed balance based on the 
\NACRE/\REACLIB{} reaction rate,
instead of consistently from the \STARLIB{} rate. This is due to limitations 
in \MESA. However, for the rates we are interested in $3\alpha$, \crate{}, 
$\carbon[12]+\carbon[12]$, and
$\oxygen[16]+\oxygen[16]$ their reverse reactions have negligible impact on the star's 
evolution. The nuclear partition
functions used to calculate the inverse rates are taken from
\citet{rauscher2000}.

We treat wind mass loss rates as in \citet{marchant:19}, where we assume 
the mass loss rate of \citet{hamann:82, hamann:95, hamann:98} with a wind efficiency of 0.1 
to account for wind clumpiness. For further discussion of the effect of wind mass loss,
see \citet{farmer:19}.

\section{Calibration of \crate{} reaction rate}\label{sec:c12_depend}

To test which temperature range we are most sensitive to, we ran models 
where we used the \STARLIB{} median
\crate{} rate, but changed the rate in one of three temperature regions 
to that of the $-3\sigma$ value (for that temperature region only).
These temperatures were chosen based on the type of helium burning 
encountered at that temperature: 
core helium burning $0.1<\rm{T}/\gk<0.3$, shell helium burning 
$0.3<\rm{T}/\gk<1.0$, and
explosive helium burning $\rm{T}>1\gk$. For the default case of the 
median \csigma, the maximum black hole mass was 46\msun{}, and for $\csigma=-3$ 
(over the whole temperature range) it was
95\msun{}. When only changing the rate during the core helium burning, 
the maximum black hole mass became 79\msun{}, for helium shell burning it was 55\msun{},
and explosive helium burning it was 46\msun{}. Thus we are most 
sensitive to changes in the \crate{} in the core helium burning temperature range,
with a smaller dependence on the shell helium burning region, and we 
are not sensitive to changes in the \crate{} reaction rate in the explosive helium burning temperature range.

The changes in the maximum black hole mass occur due to the changes 
in the carbon fraction and where those changes occur.
Changes during core helium burning primarily effect the core carbon 
fraction (see Sec. \ref{sec:results}).
The maximum black hole mass does not mass depend on the \crate{} in the
explosive helium burning regime, as no helium-rich region reaches $\rm{T}>1\gk$.

 \software{
 \texttt{mesaPlot} \citep{mesaplot},
 \texttt{mesaSDK} \citep{mesasdk},
 \texttt{ipython/jupyter} \citep{perez_2007_aa,kluyver_2016_aa},
 \texttt{matplotlib} \citep{hunter_2007_aa},
 \texttt{NumPy} \citep{der_walt_2011_aa},
 \texttt{PyMC3} \citep{salvatier16},
 \MESA \citep{paxton:11,paxton:13,paxton:15,paxton:18,paxton:19}, and 
 \texttt{pyMesa} \citep{pymesa}.
          }

\bibliographystyle{aasjournal}
\bibliography{./ppisn}

\begin{thebibliography}{}
\expandafter\ifx\csname natexlab\endcsname\relax\def\natexlab#1{#1}\fi
\providecommand{\url}[1]{\href{#1}{#1}}
\providecommand{\dodoi}[1]{doi:~\href{http://doi.org/#1}{\nolinkurl{#1}}}
\providecommand{\doeprint}[1]{\href{http://ascl.net/#1}{\nolinkurl{http://ascl.net/#1}}}
\providecommand{\doarXiv}[1]{\href{https://arxiv.org/abs/#1}{\nolinkurl{https://arxiv.org/abs/#1}}}

\bibitem[{{Abbott} {et~al.}(2018){Abbott}, {Abbott}, {Abbott}, {Abernathy},
  {Acernese}, {Ackley}, {Adams}, {Adams}, {Addesso}, {Adhikari}, {Adya},
  {Affeldt}, {Agathos}, {Agatsuma}, {Aggarwal}, {Aguiar}, {Aiello}, {Ain},
  {Ajith}, {Akutsu}, {Allen}, {Allocca}, {Altin}, {Ananyeva}, {Anderson},
  {Anderson}, {Ando}, {Appert}, {Arai}, {Araya}, {Araya}, {Areeda}, {Arnaud},
  {Arun}, {Asada}, {Ascenzi}, {Ashton}, {Aso}, {Ast}, {Aston}, {Astone},
  {Atsuta}, {Aufmuth}, {Aulbert}, {Avila-Alvarez}, \& {Awai}}]{abbott18}
{Abbott}, B.~P., {Abbott}, R., {Abbott}, T.~D., {et~al.} 2018, Living Reviews
  in Relativity, 21, 3, \dodoi{10.1007/s41114-018-0012-9}

\bibitem[{{Abbott} {et~al.}(2019{\natexlab{a}}){Abbott}, {Abbott}, {Abbott},
  {Abraham}, {Acernese}, {Ackley}, {Adams}, {Adhikari}, {Adya}, {Affeldt}, \&
  et~al.}]{ligo18b}
---. 2019{\natexlab{a}}, Physical Review X, 9, 031040,
  \dodoi{10.1103/PhysRevX.9.031040}

\bibitem[{{Abbott} {et~al.}(2019{\natexlab{b}}){Abbott}, {Abbott}, {Abbott},
  {Abraham}, {Acernese}, {Ackley}, {Adams}, {Adhikari}, {Adya}, {Affeldt}, \&
  et~al.}]{ligo18a}
---. 2019{\natexlab{b}}, \apjl, 882, L24, \dodoi{10.3847/2041-8213/ab3800}

\bibitem[{{Abbott} {et~al.}(2020{\natexlab{a}}){Abbott}, {Abbott}, {Abraham},
  {Acernese}, {Ackley}, {Adams}, {Adhikari}, {Adya}, {Affeldt}, {Agathos},
  {Agatsuma}, {Aggarwal}, {Aguiar}, {Aich}, {Aiello}, {Ain}, {Ajith}, {Akcay},
  {Allen}, {Allocca}, {Altin}, {Amato}, {Anand}, {Ananyeva}, {Anderson},
  {Anderson}, {Angelova}, {Ansoldi}, {Antier}, {Appert}, {Arai}, {Araya},
  {Areeda}, {Ar{\`e}ne}, {Arnaud}, {Aronson}, {Arun}, \& {Asali}}]{GW190521a}
{Abbott}, R., {Abbott}, T.~D., {Abraham}, S., {et~al.} 2020{\natexlab{a}},
  arXiv e-prints, arXiv:2009.01075.
\newblock \doarXiv{2009.01075}

\bibitem[{{Abbott} {et~al.}(2020{\natexlab{b}}){Abbott}, {Abbott}, {Abraham},
  {Acernese}, {Ackley}, {Adams}, {Adhikari}, {Adya}, {Affeldt}, {Agathos},
  {Agatsuma}, {Aggarwal}, {Aguiar}, {Aich}, {Aiello}, {Ain}, {Ajith}, {Akcay},
  {Allen}, {Allocca}, {Altin}, {Amato}, {Anand}, {Ananyeva}, {Anderson},
  {Anderson}, {Angelova}, {Ansoldi}, {Antier}, {Appert}, {Arai}, {Araya},
  {Areeda}, {Ar{\`e}ne}, {Arnaud}, {Aronson}, {Arun}, \& {Asali}}]{GW190521b}
---. 2020{\natexlab{b}}, arXiv e-prints, arXiv:2009.01190.
\newblock \doarXiv{2009.01190}

\bibitem[{{Acernese} {et~al.}(2015){Acernese}, {Agathos}, {Agatsuma}, {Aisa},
  {Allemandou}, {Allocca}, {Amarni}, {Astone}, {Balestri}, {Ballardin},
  {Barone}, {Baronick}, {Barsuglia}, {Basti}, {Basti}, {Bauer}, {Bavigadda},
  {Bejger}, {Beker}, {Belczynski}, {Bersanetti}, {Bertolini}, {Bitossi},
  {Bizouard}, {Bloemen}, {Blom}, {Boer}, {Bogaert}, {Bondi}, {Bondu},
  {Bonelli}, {Bonnand}, {Boschi}, {Bosi}, {Bouedo}, {Bradaschia}, {Branchesi},
  {Briant}, {Brillet}, {Brisson}, {Bulik}, {Bulten}, {Buskulic}, {Buy},
  {Cagnoli}, {Calloni}, {Campeggi}, {Canuel}, \& {Carbognani}}]{virgo15}
{Acernese}, F., {Agathos}, M., {Agatsuma}, K., {et~al.} 2015, Classical and
  Quantum Gravity, 32, 024001, \dodoi{10.1088/0264-9381/32/2/024001}

\bibitem[{{Ali-Ha{\"\i}moud} {et~al.}(2017){Ali-Ha{\"\i}moud}, {Kovetz}, \&
  {Kamionkowski}}]{alihaimoud17}
{Ali-Ha{\"\i}moud}, Y., {Kovetz}, E.~D., \& {Kamionkowski}, M. 2017, \prd, 96,
  123523, \dodoi{10.1103/PhysRevD.96.123523}

\bibitem[{{An} {et~al.}(2016){An}, {Ma}, {Fan}, {Li}, {Chen}, \& {Sun}}]{an16}
{An}, Z.-D., {Ma}, Y.-G., {Fan}, G.-T., {et~al.} 2016, \apjl, 817, L5,
  \dodoi{10.3847/2041-8205/817/1/L5}

\bibitem[{{An} {et~al.}(2015){An}, {Chen}, {Ma}, {Yu}, {Sun}, {Fan}, {Li},
  {Xu}, {Huang}, \& {Wang}}]{an15}
{An}, Z.-D., {Chen}, Z.-P., {Ma}, Y.-G., {et~al.} 2015, \prc, 92, 045802,
  \dodoi{10.1103/PhysRevC.92.045802}

\bibitem[{{Angulo} {et~al.}(1999){Angulo}, {Arnould}, {Rayet}, {Descouvemont},
  {Baye}, {Leclercq-Willain}, {Coc}, {Barhoumi}, {Aguer}, {Rolfs}, {Kunz},
  {Hammer}, {Mayer}, {Paradellis}, {Kossionides}, {Chronidou}, {Spyrou},
  {degl'Innocenti}, {Fiorentini}, {Ricci}, {Zavatarelli}, {Providencia},
  {Wolters}, {Soares}, {Grama}, {Rahighi}, {Shotter}, \& {Lamehi
  Rachti}}]{angulo99}
{Angulo}, C., {Arnould}, M., {Rayet}, M., {et~al.} 1999, \nphysa, 656, 3,
  \dodoi{10.1016/S0375-9474(99)00030-5}

\bibitem[{{Arca Sedda} {et~al.}(2020){Arca Sedda}, {Mapelli}, {Spera},
  {Benacquista}, \& {Giacobbo}}]{sedda20}
{Arca Sedda}, M., {Mapelli}, M., {Spera}, M., {Benacquista}, M., \& {Giacobbo},
  N. 2020, \apj, 894, 133, \dodoi{10.3847/1538-4357/ab88b2}

\bibitem[{{Arcavi} {et~al.}(2017){Arcavi}, {Hosseinzadeh}, {Howell}, {McCully},
  {Poznanski}, {Kasen}, {Barnes}, {Zaltzman}, {Vasylyev}, {Maoz}, \&
  {Valenti}}]{arcavi:17}
{Arcavi}, I., {Hosseinzadeh}, G., {Howell}, D.~A., {et~al.} 2017, \nat, 551,
  64, \dodoi{10.1038/nature24291}

\bibitem[{{Arnett} \& {Truran}(1969)}]{arnett69}
{Arnett}, W.~D., \& {Truran}, J.~W. 1969, \apj, 157, 339,
  \dodoi{10.1086/150072}

\bibitem[{{Barkat} {et~al.}(1967){Barkat}, {Rakavy}, \& {Sack}}]{barkat:67}
{Barkat}, Z., {Rakavy}, G., \& {Sack}, N. 1967, Physical Review Letters, 18,
  379, \dodoi{10.1103/PhysRevLett.18.379}

\bibitem[{{Batta} {et~al.}(2017){Batta}, {Ramirez-Ruiz}, \& {Fryer}}]{batta17}
{Batta}, A., {Ramirez-Ruiz}, E., \& {Fryer}, C. 2017, \apjl, 846, L15,
  \dodoi{10.3847/2041-8213/aa8506}

\bibitem[{{Belczynski} {et~al.}(2016{\natexlab{a}}){Belczynski}, {Holz},
  {Bulik}, \& {O'Shaughnessy}}]{belczynski:16nat}
{Belczynski}, K., {Holz}, D.~E., {Bulik}, T., \& {O'Shaughnessy}, R.
  2016{\natexlab{a}}, \nat, 534, 512, \dodoi{10.1038/nature18322}

\bibitem[{{Belczynski} {et~al.}(2016{\natexlab{b}}){Belczynski}, {Heger},
  {Gladysz}, {Ruiter}, {Woosley}, {Wiktorowicz}, {Chen}, {Bulik},
  {O'Shaughnessy}, {Holz}, {Fryer}, \& {Berti}}]{belczynski:16}
{Belczynski}, K., {Heger}, A., {Gladysz}, W., {et~al.} 2016{\natexlab{b}},
  \aap, 594, A97, \dodoi{10.1051/0004-6361/201628980}

\bibitem[{{Bemmerer} {et~al.}(2018){Bemmerer}, {Cowan}, {Grieger}, {Hammer},
  {Hensel}, {Junghans}, {Koppitz}, {Ludwig}, {M{\"u}ller}, {Rimarzig},
  {Reinicke}, {Schwengner}, {St{\"o}ckel}, {Sz{\"u}cs}, {Tak{\'a}cs}, {Turkat},
  {Wagner}, {Wagner}, \& {Zuber}}]{bemmerer18}
{Bemmerer}, D., {Cowan}, T.~E., {Grieger}, M., {et~al.} 2018, in European
  Physical Journal Web of Conferences, Vol. 178, European Physical Journal Web
  of Conferences, 01008, \dodoi{10.1051/epjconf/201817801008}

\bibitem[{{Bond} {et~al.}(1984){Bond}, {Arnett}, \& {Carr}}]{bond84}
{Bond}, J.~R., {Arnett}, W.~D., \& {Carr}, B.~J. 1984, \apj, 280, 825,
  \dodoi{10.1086/162057}

\bibitem[{{Boothroyd} \& {Sackmann}(1988)}]{boothroyd88}
{Boothroyd}, A.~I., \& {Sackmann}, I.~J. 1988, \apj, 328, 653,
  \dodoi{10.1086/166323}

\bibitem[{{Bouffanais} {et~al.}(2019){Bouffanais}, {Mapelli}, {Gerosa}, {Di
  Carlo}, {Giacobbo}, {Berti}, \& {Baibhav}}]{Bouffanais19}
{Bouffanais}, Y., {Mapelli}, M., {Gerosa}, D., {et~al.} 2019, \apj, 886, 25,
  \dodoi{10.3847/1538-4357/ab4a79}

\bibitem[{{Brown} {et~al.}(2001){Brown}, {Heger}, {Langer}, {Lee}, {Wellstein},
  \& {Bethe}}]{brown01}
{Brown}, G.~E., {Heger}, A., {Langer}, N., {et~al.} 2001, \na, 6, 457,
  \dodoi{10.1016/S1384-1076(01)00077-X}

\bibitem[{{Buchler} \& {Yueh}(1976)}]{Buchler1976}
{Buchler}, J.~R., \& {Yueh}, W.~R. 1976, \apj, 210, 440, \dodoi{10.1086/154847}

\bibitem[{{Burbidge} {et~al.}(1957){Burbidge}, {Burbidge}, {Fowler}, \&
  {Hoyle}}]{burbidge57}
{Burbidge}, E.~M., {Burbidge}, G.~R., {Fowler}, W.~A., \& {Hoyle}, F. 1957,
  Reviews of Modern Physics, 29, 547, \dodoi{10.1103/RevModPhys.29.547}

\bibitem[{{Carr} {et~al.}(2016){Carr}, {K{\"u}hnel}, \& {Sandstad}}]{carr:16}
{Carr}, B., {K{\"u}hnel}, F., \& {Sandstad}, M. 2016, \prd, 94, 083504,
  \dodoi{10.1103/PhysRevD.94.083504}

\bibitem[{{Cassisi} {et~al.}(2007){Cassisi}, {Potekhin}, {Pietrinferni},
  {Catelan}, \& {Salaris}}]{Cassisi2007}
{Cassisi}, S., {Potekhin}, A.~Y., {Pietrinferni}, A., {Catelan}, M., \&
  {Salaris}, M. 2007, \apj, 661, 1094, \dodoi{10.1086/516819}

\bibitem[{{Caughlan} \& {Fowler}(1988)}]{caughlan88}
{Caughlan}, G.~R., \& {Fowler}, W.~A. 1988, Atomic Data and Nuclear Data
  Tables, 40, 283, \dodoi{10.1016/0092-640X(88)90009-5}

\bibitem[{Chen {et~al.}(2014)Chen, Woosley, Heger, Almgren, \&
  Whalen}]{chen:14}
Chen, K.-J., Woosley, S., Heger, A., Almgren, A., \& Whalen, D.~J. 2014, The
  Astrophysical Journal, 792, 28.
\newblock \url{http://stacks.iop.org/0004-637X/792/i=1/a=28}

\bibitem[{{Chugunov} {et~al.}(2007){Chugunov}, {Dewitt}, \&
  {Yakovlev}}]{chugunov2007}
{Chugunov}, A.~I., {Dewitt}, H.~E., \& {Yakovlev}, D.~G. 2007, \prd, 76,
  025028, \dodoi{10.1103/PhysRevD.76.025028}

\bibitem[{{Cyburt} {et~al.}(2010){Cyburt}, {Amthor}, {Ferguson}, {Meisel},
  {Smith}, {Warren}, {Heger}, {Hoffman}, {Rauscher}, {Sakharuk}, {Schatz},
  {Thielemann}, \& {Wiescher}}]{cyburt10}
{Cyburt}, R.~H., {Amthor}, A.~M., {Ferguson}, R., {et~al.} 2010, The
  Astrophysical Journal Supplement Series, 189, 240,
  \dodoi{10.1088/0067-0049/189/1/240}

\bibitem[{{de Mink} \& {Mandel}(2016)}]{demink16}
{de Mink}, S.~E., \& {Mandel}, I. 2016, \mnras, 460, 3545,
  \dodoi{10.1093/mnras/stw1219}

\bibitem[{{deBoer} {et~al.}(2017){deBoer}, {G{\"o}rres}, {Wiescher}, {Azuma},
  {Best}, {Brune}, {Fields}, {Jones}, {Pignatari}, {Sayre}, {Smith}, {Timmes},
  \& {Uberseder}}]{deboer17}
{deBoer}, R.~J., {G{\"o}rres}, J., {Wiescher}, M., {et~al.} 2017, Reviews of
  Modern Physics, 89, 035007, \dodoi{10.1103/RevModPhys.89.035007}

\bibitem[{{Di Carlo} {et~al.}(2019){Di Carlo}, {Giacobbo}, {Mapelli},
  {Pasquato}, {Spera}, {Wang}, \& {Haardt}}]{dicarlo19}
{Di Carlo}, U.~N., {Giacobbo}, N., {Mapelli}, M., {et~al.} 2019, \mnras, 487,
  2947, \dodoi{10.1093/mnras/stz1453}

\bibitem[{{Di Carlo} {et~al.}(2020{\natexlab{a}}){Di Carlo}, {Mapelli},
  {Bouffanais}, {Giacobbo}, {Santoliquido}, {Bressan}, {Spera}, \&
  {Haardt}}]{dicarlo19b}
{Di Carlo}, U.~N., {Mapelli}, M., {Bouffanais}, Y., {et~al.}
  2020{\natexlab{a}}, \mnras, 497, 1043, \dodoi{10.1093/mnras/staa1997}

\bibitem[{{Di Carlo} {et~al.}(2020{\natexlab{b}}){Di Carlo}, {Mapelli},
  {Giacobbo}, {Spera}, {Bouffanais}, {Rastello}, {Santoliquido}, {Pasquato},
  {Ballone}, {Trani}, {Torniamenti}, \& {Haardt}}]{dicarlo20}
{Di Carlo}, U.~N., {Mapelli}, M., {Giacobbo}, N., {et~al.} 2020{\natexlab{b}},
  \mnras, \dodoi{10.1093/mnras/staa2286}

\bibitem[{{Dominik} {et~al.}(2012){Dominik}, {Belczynski}, {Fryer}, {Holz},
  {Berti}, {Bulik}, {Mandel}, \& {O'Shaughnessy}}]{dominik:12}
{Dominik}, M., {Belczynski}, K., {Fryer}, C., {et~al.} 2012, \apj, 759, 52,
  \dodoi{10.1088/0004-637X/759/1/52}

\bibitem[{{Evans} {et~al.}(2000){Evans}, {Hastings}, \&
  {Peacock}}]{evans_2000_aa}
{Evans}, M., {Hastings}, N., \& {Peacock}, B. 2000, Statistical Distributions,
  3rd edn. (Hoboken, NJ: Wiley)

\bibitem[{Farmer(2018)}]{mesaplot}
Farmer, R. 2018, rjfarmer/mesaplot, \dodoi{10.5281/zenodo.1441329}

\bibitem[{Farmer \& Bauer(2018)}]{pymesa}
Farmer, R., \& Bauer, E.~B. 2018, rjfarmer/pyMesa: Add support for 10398,
  v1.0.3,  Zenodo, \dodoi{10.5281/zenodo.1205271}

\bibitem[{{Farmer} {et~al.}(2016){Farmer}, {Fields}, {Petermann}, {Dessart},
  {Cantiello}, {Paxton}, \& {Timmes}}]{farmer16}
{Farmer}, R., {Fields}, C.~E., {Petermann}, I., {et~al.} 2016, \apjs, 227, 22,
  \dodoi{10.3847/1538-4365/227/2/22}

\bibitem[{{Farmer} {et~al.}(2015){Farmer}, {Fields}, \& {Timmes}}]{farmer15}
{Farmer}, R., {Fields}, C.~E., \& {Timmes}, F.~X. 2015, \apj, 807, 184,
  \dodoi{10.1088/0004-637X/807/2/184}

\bibitem[{{Farmer} {et~al.}(2019){Farmer}, {Renzo}, {de Mink}, {Marchant}, \&
  {Justham}}]{farmer:19}
{Farmer}, R., {Renzo}, M., {de Mink}, S.~E., {Marchant}, P., \& {Justham}, S.
  2019, \apj, 887, 53, \dodoi{10.3847/1538-4357/ab518b}

\bibitem[{{Farr} {et~al.}(2019){Farr}, {Fishbach}, {Ye}, \& {Holz}}]{farr19}
{Farr}, W.~M., {Fishbach}, M., {Ye}, J., \& {Holz}, D.~E. 2019, \apjl, 883,
  L42, \dodoi{10.3847/2041-8213/ab4284}

\bibitem[{{Ferguson} {et~al.}(2005){Ferguson}, {Alexander}, {Allard}, {Barman},
  {Bodnarik}, {Hauschildt}, {Heffner-Wong}, \& {Tamanai}}]{Ferguson2005}
{Ferguson}, J.~W., {Alexander}, D.~R., {Allard}, F., {et~al.} 2005, \apj, 623,
  585, \dodoi{10.1086/428642}

\bibitem[{{Fern{\'a}ndez} {et~al.}(2018){Fern{\'a}ndez}, {Quataert},
  {Kashiyama}, \& {Coughlin}}]{fernandez18}
{Fern{\'a}ndez}, R., {Quataert}, E., {Kashiyama}, K., \& {Coughlin}, E.~R.
  2018, \mnras, 476, 2366, \dodoi{10.1093/mnras/sty306}

\bibitem[{{Fields} {et~al.}(2016){Fields}, {Farmer}, {Petermann}, {Iliadis}, \&
  {Timmes}}]{fields16}
{Fields}, C.~E., {Farmer}, R., {Petermann}, I., {Iliadis}, C., \& {Timmes},
  F.~X. 2016, \apj, 823, 46, \dodoi{10.3847/0004-637X/823/1/46}

\bibitem[{{Fields} {et~al.}(2018){Fields}, {Timmes}, {Farmer}, {Petermann},
  {Wolf}, \& {Couch}}]{fields18}
{Fields}, C.~E., {Timmes}, F.~X., {Farmer}, R., {et~al.} 2018, The
  Astrophysical Journal Supplement Series, 234, 19,
  \dodoi{10.3847/1538-4365/aaa29b}

\bibitem[{{Fishbach} {et~al.}(2020){Fishbach}, {Farr}, \& {Holz}}]{fishbach20}
{Fishbach}, M., {Farr}, W.~M., \& {Holz}, D.~E. 2020, \apjl, 891, L31,
  \dodoi{10.3847/2041-8213/ab77c9}

\bibitem[{{Fishbach} \& {Holz}(2017)}]{fishbach:17}
{Fishbach}, M., \& {Holz}, D.~E. 2017, \apjl, 851, L25,
  \dodoi{10.3847/2041-8213/aa9bf6}

\bibitem[{{Fishbach} {et~al.}(2017){Fishbach}, {Holz}, \& {Farr}}]{fishbach17a}
{Fishbach}, M., {Holz}, D.~E., \& {Farr}, B. 2017, \apjl, 840, L24,
  \dodoi{10.3847/2041-8213/aa7045}

\bibitem[{{Fowler} \& {Hoyle}(1964)}]{fowler:64}
{Fowler}, W.~A., \& {Hoyle}, F. 1964, The Astrophysical Journal Supplement
  Series, 9, 201, \dodoi{10.1086/190103}

\bibitem[{{Fragione} \& {Kocsis}(2018)}]{fragione18}
{Fragione}, G., \& {Kocsis}, B. 2018, \prl, 121, 161103,
  \dodoi{10.1103/PhysRevLett.121.161103}

\bibitem[{{Fraley}(1968)}]{fraley:68}
{Fraley}, G.~S. 1968, \apss, 2, 96, \dodoi{10.1007/BF00651498}

\bibitem[{{Fri{\v{s}}{\v{c}}i{\'c}} {et~al.}(2019){Fri{\v{s}}{\v{c}}i{\'c}},
  {Donnelly}, \& {Milner}}]{friscic19}
{Fri{\v{s}}{\v{c}}i{\'c}}, I., {Donnelly}, T.~W., \& {Milner}, R.~G. 2019,
  \prc, 100, 025804, \dodoi{10.1103/PhysRevC.100.025804}

\bibitem[{Fruet {et~al.}(2020)Fruet, Courtin, Heine, Jenkins, Adsley, Brown,
  Canavan, Catford, Charon, Curien, Della~Negra, Duprat, Hammache, Lesrel,
  Lotay, Meyer, Montanari, Morris, Moukaddam, Nippert, Podoly\'ak, Regan,
  Ribaud, Richer, Rudigier, Shearman, de~S\'er\'eville, \& Stodel}]{fruet20}
Fruet, G., Courtin, S., Heine, M., {et~al.} 2020, Phys. Rev. Lett., 124,
  192701, \dodoi{10.1103/PhysRevLett.124.192701}

\bibitem[{{Fryer}(1999)}]{fryer99}
{Fryer}, C.~L. 1999, \apj, 522, 413, \dodoi{10.1086/307647}

\bibitem[{{Fryer} {et~al.}(2012){Fryer}, {Belczynski}, {Wiktorowicz},
  {Dominik}, {Kalogera}, \& {Holz}}]{fryer:12}
{Fryer}, C.~L., {Belczynski}, K., {Wiktorowicz}, G., {et~al.} 2012, \apj, 749,
  91, \dodoi{10.1088/0004-637X/749/1/91}

\bibitem[{{Fryer} \& {Warren}(2004)}]{fryer04}
{Fryer}, C.~L., \& {Warren}, M.~S. 2004, \apj, 601, 391, \dodoi{10.1086/380193}

\bibitem[{{Fryer} {et~al.}(2001){Fryer}, {Woosley}, \& {Heger}}]{fryer01}
{Fryer}, C.~L., {Woosley}, S.~E., \& {Heger}, A. 2001, \apj, 550, 372,
  \dodoi{10.1086/319719}

\bibitem[{{Fuller} {et~al.}(1985){Fuller}, {Fowler}, \& {Newman}}]{fuller1985}
{Fuller}, G.~M., {Fowler}, W.~A., \& {Newman}, M.~J. 1985, \apj, 293, 1,
  \dodoi{10.1086/163208}

\bibitem[{{Gerosa} \& {Berti}(2017)}]{gerosa17}
{Gerosa}, D., \& {Berti}, E. 2017, \prd, 95, 124046,
  \dodoi{10.1103/PhysRevD.95.124046}

\bibitem[{{Gerosa} \& {Berti}(2019)}]{gerosa19}
---. 2019, \prd, 100, 041301, \dodoi{10.1103/PhysRevD.100.041301}

\bibitem[{{Gilkis} {et~al.}(2016){Gilkis}, {Soker}, \& {Papish}}]{gilkis16}
{Gilkis}, A., {Soker}, N., \& {Papish}, O. 2016, \apj, 826, 178,
  \dodoi{10.3847/0004-637X/826/2/178}

\bibitem[{{Gomez} {et~al.}(2019){Gomez}, {Berger}, {Nicholl}, {Blanchard},
  {Villar}, {Patton}, {Chornock}, {Leja}, {Hosseinzadeh}, \&
  {Cowperthwaite}}]{gomez19}
{Gomez}, S., {Berger}, E., {Nicholl}, M., {et~al.} 2019, \apj, 881, 87,
  \dodoi{10.3847/1538-4357/ab2f92}

\bibitem[{{Graham} {et~al.}(2020){Graham}, {Ford}, {McKernan}, {Ross}, {Stern},
  {Burdge}, {Coughlin}, {Djorgovski}, {Drake}, {Duev}, {Kasliwal}, {Mahabal},
  {van Velzen}, {Belecki}, {Bellm}, {Burruss}, {Cenko}, {Cunningham}, {Helou},
  {Kulkarni}, {Masci}, {Prince}, {Reiley}, {Rodriguez}, {Rusholme}, {Smith}, \&
  {Soumagnac}}]{graham20}
{Graham}, M.~J., {Ford}, K.~E.~S., {McKernan}, B., {et~al.} 2020, \prl, 124,
  251102, \dodoi{10.1103/PhysRevLett.124.251102}

\bibitem[{{Hamann} \& {Koesterke}(1998)}]{hamann:98}
{Hamann}, W.~R., \& {Koesterke}, L. 1998, \aap, 335, 1003

\bibitem[{{Hamann} {et~al.}(1995){Hamann}, {Koesterke}, \&
  {Wessolowski}}]{hamann:95}
{Hamann}, W.-R., {Koesterke}, L., \& {Wessolowski}, U. 1995, \aap, 299, 151

\bibitem[{{Hamann} {et~al.}(1982){Hamann}, {Schoenberner}, \&
  {Heber}}]{hamann:82}
{Hamann}, W.-R., {Schoenberner}, D., \& {Heber}, U. 1982, \aap, 116, 273

\bibitem[{{Hammache} {et~al.}(2016){Hammache}, {Oulebsir}, {Roussel},
  {Pellegriti}, {Audouin}, {Beaumel}, {Bouda}, {Descouvemont}, {Fortier},
  {Gaudefroy}, {Kiener}, {Lefebvre-Schuhl}, \& {Tatischeff}}]{hammache16}
{Hammache}, F., {Oulebsir}, N., {Roussel}, P., {et~al.} 2016, in Journal of
  Physics Conference Series, Vol. 665, Journal of Physics Conference Series,
  012007, \dodoi{10.1088/1742-6596/665/1/012007}

\bibitem[{{Hammer} {et~al.}(2005){Hammer}, {Fey}, {Kunz}, {Kiener},
  {Tatischeff}, {Haas}, {Weil}, {Assun{\c{c}}{\~a}o}, {Beck},
  {Boukari-Pelissie}, {Coc}, {Correia}, {Courtin}, {Fleurot}, {Galanopoulos},
  {Grama}, {Hammache}, {Harissopulos}, {Korichi}, {Krmpoti{\'c}}, {Le Du},
  {Lopez-Martens}, {Malcherek}, {Meunier}, {Papka}, {Paradellis}, {Rousseau},
  {Rowley}, {Staudt}, \& {Szilner}}]{hammer02}
{Hammer}, J.~W., {Fey}, M., {Kunz}, R., {et~al.} 2005, \nphysa, 752, 514,
  \dodoi{10.1016/j.nuclphysa.2005.02.056}

\bibitem[{{Heger} {et~al.}(2003){Heger}, {Fryer}, {Woosley}, {Langer}, \&
  {Hartmann}}]{heger:03}
{Heger}, A., {Fryer}, C.~L., {Woosley}, S.~E., {Langer}, N., \& {Hartmann},
  D.~H. 2003, \apj, 591, 288, \dodoi{10.1086/375341}

\bibitem[{{Heger} \& {Woosley}(2002)}]{heger02}
{Heger}, A., \& {Woosley}, S.~E. 2002, \apj, 567, 532, \dodoi{10.1086/338487}

\bibitem[{{Heger} {et~al.}(2002){Heger}, {Woosley}, {Rauscher}, {Hoffman}, \&
  {Boyes}}]{heger02b}
{Heger}, A., {Woosley}, S.~E., {Rauscher}, T., {Hoffman}, R.~D., \& {Boyes},
  M.~M. 2002, \nar, 46, 463, \dodoi{10.1016/S1387-6473(02)00184-7}

\bibitem[{{Heger} {et~al.}(2005){Heger}, {Woosley}, \& {Spruit}}]{heger05}
{Heger}, A., {Woosley}, S.~E., \& {Spruit}, H.~C. 2005, \apj, 626, 350,
  \dodoi{10.1086/429868}

\bibitem[{{Hoffman} {et~al.}(1999){Hoffman}, {Woosley}, {Weaver}, {Rauscher},
  \& {Thielemann}}]{hoffman99}
{Hoffman}, R.~D., {Woosley}, S.~E., {Weaver}, T.~A., {Rauscher}, T., \&
  {Thielemann}, F.~K. 1999, \apj, 521, 735, \dodoi{10.1086/307568}

\bibitem[{{Holt} {et~al.}(2018){Holt}, {Filippone}, \& {Pieper}}]{holt18}
{Holt}, R.~J., {Filippone}, B.~W., \& {Pieper}, S.~C. 2018, arXiv e-prints,
  arXiv:1809.10176.
\newblock \doarXiv{1809.10176}

\bibitem[{{Holt} {et~al.}(2019){Holt}, {Filippone}, \& {Pieper}}]{holt19}
---. 2019, \prc, 99, 055802, \dodoi{10.1103/PhysRevC.99.055802}

\bibitem[{{Holz} \& {Hughes}(2005)}]{holz15}
{Holz}, D.~E., \& {Hughes}, S.~A. 2005, \apj, 629, 15, \dodoi{10.1086/431341}

\bibitem[{{Hummel} {et~al.}(2012){Hummel}, {Pawlik}, {Milosavljevi{\'c}}, \&
  {Bromm}}]{hummel12}
{Hummel}, J.~A., {Pawlik}, A.~H., {Milosavljevi{\'c}}, M., \& {Bromm}, V. 2012,
  \apj, 755, 72, \dodoi{10.1088/0004-637X/755/1/72}

\bibitem[{Hunter(2007)}]{hunter_2007_aa}
Hunter, J.~D. 2007, Computing In Science \&amp; Engineering, 9, 90

\bibitem[{{Iglesias} \& {Rogers}(1993)}]{Iglesias1993}
{Iglesias}, C.~A., \& {Rogers}, F.~J. 1993, \apj, 412, 752,
  \dodoi{10.1086/172958}

\bibitem[{{Iglesias} \& {Rogers}(1996)}]{Iglesias1996}
---. 1996, \apj, 464, 943, \dodoi{10.1086/177381}

\bibitem[{{Iliadis} {et~al.}(2002){Iliadis}, {Champagne}, {Jos{\'e}},
  {Starrfield}, \& {Tupper}}]{iliadis02}
{Iliadis}, C., {Champagne}, A., {Jos{\'e}}, J., {Starrfield}, S., \& {Tupper},
  P. 2002, \apjs, 142, 105, \dodoi{10.1086/341400}

\bibitem[{{Iliadis} {et~al.}(2010{\natexlab{a}}){Iliadis}, {Longland},
  {Champagne}, \& {Coc}}]{iliadis_2010_ab}
{Iliadis}, C., {Longland}, R., {Champagne}, A.~E., \& {Coc}, A.
  2010{\natexlab{a}}, Nuclear Physics A, 841, 251,
  \dodoi{10.1016/j.nuclphysa.2010.04.010}

\bibitem[{{Iliadis} {et~al.}(2010{\natexlab{b}}){Iliadis}, {Longland},
  {Champagne}, {Coc}, \& {Fitzgerald}}]{iliadis_2010_aa}
{Iliadis}, C., {Longland}, R., {Champagne}, A.~E., {Coc}, A., \& {Fitzgerald},
  R. 2010{\natexlab{b}}, Nuclear Physics A, 841, 31,
  \dodoi{10.1016/j.nuclphysa.2010.04.009}

\bibitem[{{Itoh} {et~al.}(1996){Itoh}, {Hayashi}, {Nishikawa}, \&
  {Kohyama}}]{itoh96}
{Itoh}, N., {Hayashi}, H., {Nishikawa}, A., \& {Kohyama}, Y. 1996, The
  Astrophysical Journal Supplement Series, 102, 411, \dodoi{10.1086/192264}

\bibitem[{{Kimball} {et~al.}(2020){Kimball}, {Talbot}, {Berry}, {Carney},
  {Zevin}, {Thrane}, \& {Kalogera}}]{kimball20}
{Kimball}, C., {Talbot}, C., {Berry}, C. P.~L., {et~al.} 2020, arXiv e-prints,
  arXiv:2005.00023.
\newblock \doarXiv{2005.00023}

\bibitem[{Kluyver {et~al.}(2016)Kluyver, Ragan-Kelley, P{\'e}rez, Granger,
  Bussonnier, Frederic, Kelley, Hamrick, Grout, Corlay,
  {et~al.}}]{kluyver_2016_aa}
Kluyver, T., Ragan-Kelley, B., P{\'e}rez, F., {et~al.} 2016, in Positioning and
  Power in Academic Publishing: Players, Agents and Agendas: Proceedings of the
  20th International Conference on Electronic Publishing, IOS Press, 87

\bibitem[{{Kulkarni} {et~al.}(1993){Kulkarni}, {Hut}, \&
  {McMillan}}]{Kulkarni93}
{Kulkarni}, S.~R., {Hut}, P., \& {McMillan}, S. 1993, \nat, 364, 421,
  \dodoi{10.1038/364421a0}

\bibitem[{{Kunz} {et~al.}(2002){Kunz}, {Fey}, {Jaeger}, {Mayer}, {Hammer},
  {Staudt}, {Harissopulos}, \& {Paradellis}}]{kunz02}
{Kunz}, R., {Fey}, M., {Jaeger}, M., {et~al.} 2002, \apj, 567, 643,
  \dodoi{10.1086/338384}

\bibitem[{{Langanke} \& {Mart{\'{\i}}nez-Pinedo}(2000)}]{langanke2000}
{Langanke}, K., \& {Mart{\'{\i}}nez-Pinedo}, G. 2000, Nuclear Physics A, 673,
  481, \dodoi{10.1016/S0375-9474(00)00131-7}

\bibitem[{{Leung} {et~al.}(2019){Leung}, {Nomoto}, \& {Blinnikov}}]{leung19}
{Leung}, S.-C., {Nomoto}, K., \& {Blinnikov}, S. 2019, \apj, 887, 72,
  \dodoi{10.3847/1538-4357/ab4fe5}

\bibitem[{{LIGO Scientific Collaboration} {et~al.}(2015){LIGO Scientific
  Collaboration}, {Aasi}, {Abbott}, {Abbott}, {Abbott}, {Abernathy}, {Ackley},
  {Adams}, {Adams}, {Addesso}, {Adhikari}, {Adya}, {Affeldt}, {Aggarwal},
  {Aguiar}, {Ain}, {Ajith}, {Alemic}, {Allen}, {Amariutei}, {Anderson},
  {Anderson}, {Arai}, {Araya}, {Arceneaux}, {Areeda}, {Ashton}, {Ast}, {Aston},
  {Aufmuth}, {Aulbert}, {Aylott}, {Babak}, {Baker}, {Ballmer}, {Barayoga},
  {Barbet}, {Barclay}, {Barish}, {Barker}, {Barr}, {Barsotti}, {Bartlett},
  {Barton}, \& {Bartos}}]{ligo15a}
{LIGO Scientific Collaboration}, {Aasi}, J., {Abbott}, B.~P., {et~al.} 2015,
  Classical and Quantum Gravity, 32, 074001,
  \dodoi{10.1088/0264-9381/32/7/074001}

\bibitem[{{Lippuner} \& {Roberts}(2017)}]{lippuner17}
{Lippuner}, J., \& {Roberts}, L.~F. 2017, \apjs, 233, 18,
  \dodoi{10.3847/1538-4365/aa94cb}

\bibitem[{{Longland} {et~al.}(2010){Longland}, {Iliadis}, {Champagne},
  {Newton}, {Ugalde}, {Coc}, \& {Fitzgerald}}]{longland_2010_aa}
{Longland}, R., {Iliadis}, C., {Champagne}, A.~E., {et~al.} 2010, Nuclear
  Physics A, 841, 1, \dodoi{10.1016/j.nuclphysa.2010.04.008}

\bibitem[{{Lovegrove} \& {Woosley}(2013)}]{lovegrove:13}
{Lovegrove}, E., \& {Woosley}, S.~E. 2013, \apj, 769, 109,
  \dodoi{10.1088/0004-637X/769/2/109}

\bibitem[{{Lunnan} {et~al.}(2018){Lunnan}, {Fransson}, {Vreeswijk}, {Woosley},
  {Leloudas}, {Perley}, {Quimby}, {Yan}, {Blagorodnova}, {Bue}, {Cenko}, {De
  Cia}, {Cook}, {Fremling}, {Gatkine}, {Gal-Yam}, {Kasliwal}, {Kulkarni},
  {Masci}, {Nugent}, {Nyholm}, {Rubin}, {Suzuki}, \& {Wozniak}}]{lunnan:18}
{Lunnan}, R., {Fransson}, C., {Vreeswijk}, P.~M., {et~al.} 2018, Nature
  Astronomy, \dodoi{10.1038/s41550-018-0568-z}

\bibitem[{{Mandel}(2010)}]{mandel10}
{Mandel}, I. 2010, \prd, 81, 084029, \dodoi{10.1103/PhysRevD.81.084029}

\bibitem[{{Mandel} \& {de Mink}(2016)}]{mandel:16b}
{Mandel}, I., \& {de Mink}, S.~E. 2016, \mnras, 458, 2634,
  \dodoi{10.1093/mnras/stw379}

\bibitem[{{Mandel} {et~al.}(2019){Mandel}, {Farr}, \& {Gair}}]{mandel19}
{Mandel}, I., {Farr}, W.~M., \& {Gair}, J.~R. 2019, \mnras, 486, 1086,
  \dodoi{10.1093/mnras/stz896}

\bibitem[{{Mangiagli} {et~al.}(2019){Mangiagli}, {Bonetti}, {Sesana}, \&
  {Colpi}}]{mangiagli19}
{Mangiagli}, A., {Bonetti}, M., {Sesana}, A., \& {Colpi}, M. 2019, \apjl, 883,
  L27, \dodoi{10.3847/2041-8213/ab3f33}

\bibitem[{{Marchant} {et~al.}(2016){Marchant}, {Langer}, {Podsiadlowski},
  {Tauris}, \& {Moriya}}]{marchant:16}
{Marchant}, P., {Langer}, N., {Podsiadlowski}, P., {Tauris}, T.~M., \&
  {Moriya}, T.~J. 2016, \aap, 588, A50, \dodoi{10.1051/0004-6361/201628133}

\bibitem[{{Marchant} {et~al.}(2019){Marchant}, {Renzo}, {Farmer}, {Pappas},
  {Taam}, {de Mink}, \& {Kalogera}}]{marchant:19}
{Marchant}, P., {Renzo}, M., {Farmer}, R., {et~al.} 2019, \apj, 882, 36,
  \dodoi{10.3847/1538-4357/ab3426}

\bibitem[{{McKernan} {et~al.}(2018){McKernan}, {Ford}, {Bellovary}, {Leigh},
  {Haiman}, {Kocsis}, {Lyra}, {Mac Low}, {Metzger}, {O'Dowd}, {Endlich}, \&
  {Rosen}}]{McKernan18}
{McKernan}, B., {Ford}, K.~E.~S., {Bellovary}, J., {et~al.} 2018, \apj, 866,
  66, \dodoi{10.3847/1538-4357/aadae5}

\bibitem[{{Metcalfe}(2003)}]{metcalfe03}
{Metcalfe}, T.~S. 2003, \apjl, 587, L43, \dodoi{10.1086/375044}

\bibitem[{{Metcalfe} {et~al.}(2002){Metcalfe}, {Salaris}, \&
  {Winget}}]{metcalfe02}
{Metcalfe}, T.~S., {Salaris}, M., \& {Winget}, D.~E. 2002, \apj, 573, 803,
  \dodoi{10.1086/340796}

\bibitem[{{Metcalfe} {et~al.}(2001){Metcalfe}, {Winget}, \&
  {Charbonneau}}]{metcalfe01}
{Metcalfe}, T.~S., {Winget}, D.~E., \& {Charbonneau}, P. 2001, \apj, 557, 1021,
  \dodoi{10.1086/321643}

\bibitem[{{Nadezhin}(1980)}]{nadezhin:80}
{Nadezhin}, D.~K. 1980, \apss, 69, 115, \dodoi{10.1007/BF00638971}

\bibitem[{{Nicholl} {et~al.}(2020){Nicholl}, {Blanchard}, {Berger}, {Chornock},
  {Margutti}, {Gomez}, {Lunnan}, {Miller}, {Fong}, {Terreran},
  {Vigna-G{\'o}mez}, {Bhirombhakdi}, {Bieryla}, {Challis}, {Laher}, {Masci}, \&
  {Paterson}}]{nicholl20}
{Nicholl}, M., {Blanchard}, P.~K., {Berger}, E., {et~al.} 2020, Nature
  Astronomy, \dodoi{10.1038/s41550-020-1066-7}

\bibitem[{Nitz {et~al.}(2020)Nitz, Dent, Davies, Kumar, Capano, Harry, Mozzon,
  Nuttall, Lundgren, \& T{\'{a}}pai}]{Nitz20}
Nitz, A.~H., Dent, T., Davies, G.~S., {et~al.} 2020, The Astrophysical Journal,
  891, 123, \dodoi{10.3847/1538-4357/ab733f}

\bibitem[{{Oda} {et~al.}(1994){Oda}, {Hino}, {Muto}, {Takahara}, \&
  {Sato}}]{oda1994}
{Oda}, T., {Hino}, M., {Muto}, K., {Takahara}, M., \& {Sato}, K. 1994, Atomic
  Data and Nuclear Data Tables, 56, 231, \dodoi{10.1006/adnd.1994.1007}

\bibitem[{{Pastorello} {et~al.}(2007){Pastorello}, {Smartt}, {Mattila},
  {Eldridge}, {Young}, {Itagaki}, {Yamaoka}, {Navasardyan}, {Valenti}, {Patat},
  {Agnoletto}, {Augusteijn}, {Benetti}, {Cappellaro}, {Boles}, {Bonnet-Bidaud},
  {Botticella}, {Bufano}, {Cao}, {Deng}, {Dennefeld}, {Elias-Rosa},
  {Harutyunyan}, {Keenan}, {Iijima}, {Lorenzi}, {Mazzali}, {Meng}, {Nakano},
  {Nielsen}, {Smoker}, {Stanishev}, {Turatto}, {Xu}, \&
  {Zampieri}}]{pastorello07}
{Pastorello}, A., {Smartt}, S.~J., {Mattila}, S., {et~al.} 2007, \nat, 447,
  829, \dodoi{10.1038/nature05825}

\bibitem[{{Paxton} {et~al.}(2011){Paxton}, {Bildsten}, {Dotter}, {Herwig},
  {Lesaffre}, \& {Timmes}}]{paxton:11}
{Paxton}, B., {Bildsten}, L., {Dotter}, A., {et~al.} 2011, \apjs, 192, 3,
  \dodoi{10.1088/0067-0049/192/1/3}

\bibitem[{{Paxton} {et~al.}(2013){Paxton}, {Cantiello}, {Arras}, {Bildsten},
  {Brown}, {Dotter}, {Mankovich}, {Montgomery}, {Stello}, {Timmes}, \&
  {Townsend}}]{paxton:13}
{Paxton}, B., {Cantiello}, M., {Arras}, P., {et~al.} 2013, \apjs, 208, 4,
  \dodoi{10.1088/0067-0049/208/1/4}

\bibitem[{{Paxton} {et~al.}(2015){Paxton}, {Marchant}, {Schwab}, {Bauer},
  {Bildsten}, {Cantiello}, {Dessart}, {Farmer}, {Hu}, {Langer}, {Townsend},
  {Townsley}, \& {Timmes}}]{paxton:15}
{Paxton}, B., {Marchant}, P., {Schwab}, J., {et~al.} 2015, \apjs, 220, 15,
  \dodoi{10.1088/0067-0049/220/1/15}

\bibitem[{{Paxton} {et~al.}(2018){Paxton}, {Schwab}, {Bauer}, {Bildsten},
  {Blinnikov}, {Duffell}, {Farmer}, {Goldberg}, {Marchant}, {Sorokina},
  {Thoul}, {Townsend}, \& {Timmes}}]{paxton:18}
{Paxton}, B., {Schwab}, J., {Bauer}, E.~B., {et~al.} 2018, \apjs, 234, 34,
  \dodoi{10.3847/1538-4365/aaa5a8}

\bibitem[{{Paxton} {et~al.}(2019){Paxton}, {Smolec}, {Schwab}, {Gautschy},
  {Bildsten}, {Cantiello}, {Dotter}, {Farmer}, {Goldberg}, {Jermyn}, {Kanbur},
  {Marchant}, {Thoul}, {Townsend}, {Wolf}, {Zhang}, \& {Timmes}}]{paxton:19}
{Paxton}, B., {Smolec}, R., {Schwab}, J., {et~al.} 2019, \apjs, 243, 10,
  \dodoi{10.3847/1538-4365/ab2241}

\bibitem[{P{\'e}rez \& Granger(2007)}]{perez_2007_aa}
P{\'e}rez, F., \& Granger, B.~E. 2007, Computing in Science \& Engineering, 9,
  21

\bibitem[{{Pols} {et~al.}(1995){Pols}, {Tout}, {Eggleton}, \& {Han}}]{Pols1995}
{Pols}, O.~R., {Tout}, C.~A., {Eggleton}, P.~P., \& {Han}, Z. 1995, \mnras,
  274, 964, \dodoi{10.1093/mnras/274.3.964}

\bibitem[{{Portegies Zwart} \& {McMillan}(2000)}]{Portegies00}
{Portegies Zwart}, S.~F., \& {McMillan}, S. L.~W. 2000, \apj, 528, L17,
  \dodoi{10.1086/312422}

\bibitem[{{Potekhin} \& {Chabrier}(2010)}]{Potekhin2010}
{Potekhin}, A.~Y., \& {Chabrier}, G. 2010, Contributions to Plasma Physics, 50,
  82, \dodoi{10.1002/ctpp.201010017}

\bibitem[{{Quataert} {et~al.}(2019){Quataert}, {Lecoanet}, \&
  {Coughlin}}]{quataert19}
{Quataert}, E., {Lecoanet}, D., \& {Coughlin}, E.~R. 2019, \mnras, 485, L83,
  \dodoi{10.1093/mnrasl/slz031}

\bibitem[{{Rakavy} \& {Shaviv}(1967)}]{rakavy:67}
{Rakavy}, G., \& {Shaviv}, G. 1967, \apj, 148, 803, \dodoi{10.1086/149204}

\bibitem[{{Rauscher} {et~al.}(2016){Rauscher}, {Nishimura}, {Hirschi},
  {Cescutti}, {Murphy}, \& {Heger}}]{rauscher16}
{Rauscher}, T., {Nishimura}, N., {Hirschi}, R., {et~al.} 2016, \mnras, 463,
  4153, \dodoi{10.1093/mnras/stw2266}

\bibitem[{{Rauscher} \& {Thielemann}(2000)}]{rauscher2000}
{Rauscher}, T., \& {Thielemann}, F.-K. 2000, Atomic Data and Nuclear Data
  Tables, 75, 1, \dodoi{10.1006/adnd.2000.0834}

\bibitem[{{Reg{\H{o}}s} {et~al.}(2020){Reg{\H{o}}s}, {Vink{\'o}}, \&
  {Ziegler}}]{regos20}
{Reg{\H{o}}s}, E., {Vink{\'o}}, J., \& {Ziegler}, B.~L. 2020, \apj, 894, 94,
  \dodoi{10.3847/1538-4357/ab8636}

\bibitem[{{Renzo} {et~al.}(2020{\natexlab{a}}){Renzo}, {Farmer}, {Justham},
  {G{\"o}tberg}, {de Mink}, {Zapartas}, {Marchant}, \& {Smith}}]{renzo20csm}
{Renzo}, M., {Farmer}, R., {Justham}, S., {et~al.} 2020{\natexlab{a}}, \aap,
  640, A56, \dodoi{10.1051/0004-6361/202037710}

\bibitem[{{Renzo} {et~al.}(2020{\natexlab{b}}){Renzo}, {Farmer}, {Justham}, {de
  Mink}, {G{\"o}tberg}, \& {Marchant}}]{renzo20conv}
{Renzo}, M., {Farmer}, R.~J., {Justham}, S., {et~al.} 2020{\natexlab{b}},
  \mnras, 493, 4333, \dodoi{10.1093/mnras/staa549}

\bibitem[{{Rockefeller} {et~al.}(2006){Rockefeller}, {Fryer}, \&
  {Li}}]{rockefeller:06}
{Rockefeller}, G., {Fryer}, C.~L., \& {Li}, H. 2006, arXiv e-prints, astro.
\newblock \doarXiv{astro-ph/0608028}

\bibitem[{{Rodriguez} {et~al.}(2018){Rodriguez}, {Amaro-Seoane}, {Chatterjee},
  {Kremer}, {Rasio}, {Samsing}, {Ye}, \& {Zevin}}]{Rodriguez18}
{Rodriguez}, C.~L., {Amaro-Seoane}, P., {Chatterjee}, S., {et~al.} 2018, \prd,
  98, 123005, \dodoi{10.1103/PhysRevD.98.123005}

\bibitem[{{Rodriguez} {et~al.}(2016){Rodriguez}, {Chatterjee}, \&
  {Rasio}}]{rodriguez16a}
{Rodriguez}, C.~L., {Chatterjee}, S., \& {Rasio}, F.~A. 2016, \prd, 93, 084029,
  \dodoi{10.1103/PhysRevD.93.084029}

\bibitem[{{Rodriguez} {et~al.}(2019){Rodriguez}, {Zevin}, {Amaro-Seoane},
  {Chatterjee}, {Kremer}, {Rasio}, \& {Ye}}]{rodriguez19}
{Rodriguez}, C.~L., {Zevin}, M., {Amaro-Seoane}, P., {et~al.} 2019, \prd, 100,
  043027, \dodoi{10.1103/PhysRevD.100.043027}

\bibitem[{{Rogers} \& {Nayfonov}(2002)}]{Rogers2002}
{Rogers}, F.~J., \& {Nayfonov}, A. 2002, \apj, 576, 1064,
  \dodoi{10.1086/341894}

\bibitem[{{Roupas} \& {Kazanas}(2019)}]{roupas19}
{Roupas}, Z., \& {Kazanas}, D. 2019, \aap, 632, L8,
  \dodoi{10.1051/0004-6361/201937002}

\bibitem[{{Salaris} {et~al.}(1997){Salaris}, {Dom{\'\i}nguez},
  {Garc{\'\i}a-Berro}, {Hernanz}, {Isern}, \& {Mochkovitch}}]{salaris97}
{Salaris}, M., {Dom{\'\i}nguez}, I., {Garc{\'\i}a-Berro}, E., {et~al.} 1997,
  \apj, 486, 413, \dodoi{10.1086/304483}

\bibitem[{{Sallaska} {et~al.}(2013){Sallaska}, {Iliadis}, {Champange},
  {Goriely}, {Starrfield}, \& {Timmes}}]{sallaska13}
{Sallaska}, A.~L., {Iliadis}, C., {Champange}, A.~E., {et~al.} 2013, The
  Astrophysical Journal Supplement Series, 207, 18,
  \dodoi{10.1088/0067-0049/207/1/18}

\bibitem[{{Salvatier} {et~al.}(2016){Salvatier}, {Wiecki{\^a}}, \&
  {Fonnesbeck}}]{salvatier16}
{Salvatier}, J., {Wiecki{\^a}}, T.~V., \& {Fonnesbeck}, C. 2016, {PyMC3: Python
  probabilistic programming framework}.
\newblock \doeprint{1610.016}

\bibitem[{{Saumon} {et~al.}(1995){Saumon}, {Chabrier}, \& {van
  Horn}}]{Saumon1995}
{Saumon}, D., {Chabrier}, G., \& {van Horn}, H.~M. 1995, \apjs, 99, 713,
  \dodoi{10.1086/192204}

\bibitem[{{Schutz}(1986)}]{schutz:86}
{Schutz}, B.~F. 1986, \nat, 323, 310, \dodoi{10.1038/323310a0}

\bibitem[{Shen {et~al.}(2020)Shen, Guo, deBoer, Li, Li, Tang, Pang, Adhikari,
  Basu, Su, Yan, Fan, Liu, Chen, Han, Li, Lian, Ma, Nan, Nan, Wang, Zeng,
  Zhang, \& Liu}]{shen20}
Shen, Y.~P., Guo, B., deBoer, R.~J., {et~al.} 2020, Phys. Rev. Lett., 124,
  162701, \dodoi{10.1103/PhysRevLett.124.162701}

\bibitem[{{Spera} {et~al.}(2019){Spera}, {Mapelli}, {Giacobbo}, {Trani},
  {Bressan}, \& {Costa}}]{spera19}
{Spera}, M., {Mapelli}, M., {Giacobbo}, N., {et~al.} 2019, \mnras, 485, 889,
  \dodoi{10.1093/mnras/stz359}

\bibitem[{{Stone} {et~al.}(2017){Stone}, {Metzger}, \& {Haiman}}]{stone:17}
{Stone}, N.~C., {Metzger}, B.~D., \& {Haiman}, Z. 2017, \mnras, 464, 946,
  \dodoi{10.1093/mnras/stw2260}

\bibitem[{{Stothers}(1999)}]{stothers:99}
{Stothers}, R.~B. 1999, \mnras, 305, 365,
  \dodoi{10.1046/j.1365-8711.1999.02444.x}

\bibitem[{{Straniero} {et~al.}(2003){Straniero}, {Dom{\'\i}nguez}, {Imbriani},
  \& {Piersanti}}]{straniero03}
{Straniero}, O., {Dom{\'\i}nguez}, I., {Imbriani}, G., \& {Piersanti}, L. 2003,
  \apj, 583, 878, \dodoi{10.1086/345427}

\bibitem[{{Sukhbold} \& {Adams}(2020)}]{sukhbold20}
{Sukhbold}, T., \& {Adams}, S. 2020, \mnras, 492, 2578,
  \dodoi{10.1093/mnras/staa059}

\bibitem[{{Takahashi}(2018)}]{takahashi18}
{Takahashi}, K. 2018, \apj, 863, 153, \dodoi{10.3847/1538-4357/aad2d2}

\bibitem[{Tan {et~al.}(2020)Tan, Boeltzig, Dulal, deBoer, Frentz, Henderson,
  Howard, Kelmar, Kolata, Long, Macon, Moylan, Peaslee, Renaud, Seymour,
  Seymour, Vande~Kolk, Wiescher, Aguilera, Amador-Valenzuela, Lizcano, \&
  Martinez-Quiroz}]{tan20}
Tan, W.~P., Boeltzig, A., Dulal, C., {et~al.} 2020, Phys. Rev. Lett., 124,
  192702, \dodoi{10.1103/PhysRevLett.124.192702}

\bibitem[{{Thielemann} {et~al.}(1996){Thielemann}, {Nomoto}, \&
  {Hashimoto}}]{thielemann96}
{Thielemann}, F.-K., {Nomoto}, K., \& {Hashimoto}, M.-A. 1996, \apj, 460, 408,
  \dodoi{10.1086/176980}

\bibitem[{{Timmes} \& {Swesty}(2000)}]{Timmes2000}
{Timmes}, F.~X., \& {Swesty}, F.~D. 2000, \apjs, 126, 501,
  \dodoi{10.1086/313304}

\bibitem[{{Toro} {et~al.}(1994){Toro}, {Spruce}, \& {Speares}}]{Toro1994}
{Toro}, E.~F., {Spruce}, M., \& {Speares}, W. 1994, Shock Waves, 4, 25,
  \dodoi{10.1007/BF01414629}

\bibitem[{Townsend(2019)}]{mesasdk}
Townsend, R. 2019, MESA SDK for Linux: 20190911, \dodoi{10.5281/zenodo.2603170}

\bibitem[{{Tumino} {et~al.}(2018){Tumino}, {Spitaleri}, {La Cognata},
  {Cherubini}, {Guardo}, {Gulino}, {Hayakawa}, {Indelicato}, {Lamia},
  {Petrascu}, {Pizzone}, {Puglia}, {Rapisarda}, {Romano}, {Sergi},
  {Spart{\'a}}, \& {Trache}}]{tumino18}
{Tumino}, A., {Spitaleri}, C., {La Cognata}, M., {et~al.} 2018, \nat, 557, 687,
  \dodoi{10.1038/s41586-018-0149-4}

\bibitem[{{Tur} {et~al.}(2007){Tur}, {Heger}, \& {Austin}}]{tur07}
{Tur}, C., {Heger}, A., \& {Austin}, S.~M. 2007, \apj, 671, 821,
  \dodoi{10.1086/523095}

\bibitem[{{Tutukov} \& {Yungelson}(1993)}]{tutukov93}
{Tutukov}, A.~V., \& {Yungelson}, L.~R. 1993, \mnras, 260, 675,
  \dodoi{10.1093/mnras/260.3.675}

\bibitem[{{Udall} {et~al.}(2019){Udall}, {Jani}, {Lange}, {O'Shaughnessy},
  {Clark}, {Cadonati}, {Shoemaker}, \& {Holley-Bockelmann}}]{udall19}
{Udall}, R., {Jani}, K., {Lange}, J., {et~al.} 2019, arXiv e-prints,
  arXiv:1912.10533.
\newblock \doarXiv{1912.10533}

\bibitem[{van~der Walt {et~al.}(2011)van~der Walt, Colbert, \&
  Varoquaux}]{der_walt_2011_aa}
van~der Walt, S., Colbert, S.~C., \& Varoquaux, G. 2011, Computing in Science
  Engineering, 13, 22, \dodoi{10.1109/MCSE.2011.37}

\bibitem[{{van Son} {et~al.}(2020){van Son}, {De Mink}, {Broekgaarden},
  {Renzo}, {Justham}, {Laplace}, {Mor{\'a}n-Fraile}, {Hendriks}, \&
  {Farmer}}]{vanson20}
{van Son}, L.~A.~C., {De Mink}, S.~E., {Broekgaarden}, F.~S., {et~al.} 2020,
  \apj, 897, 100, \dodoi{10.3847/1538-4357/ab9809}

\bibitem[{{Vigna-G{\'o}mez} {et~al.}(2019){Vigna-G{\'o}mez}, {Justham},
  {Mandel}, {de Mink}, \& {Podsiadlowski}}]{vigna19}
{Vigna-G{\'o}mez}, A., {Justham}, S., {Mandel}, I., {de Mink}, S.~E., \&
  {Podsiadlowski}, P. 2019, \apjl, 876, L29, \dodoi{10.3847/2041-8213/ab1bdf}

\bibitem[{{Villar} {et~al.}(2018){Villar}, {Nicholl}, \& {Berger}}]{villar18}
{Villar}, V.~A., {Nicholl}, M., \& {Berger}, E. 2018, \apj, 869, 166,
  \dodoi{10.3847/1538-4357/aaee6a}

\bibitem[{{Vitale} {et~al.}(2017){Vitale}, {Lynch}, {Raymond}, {Sturani},
  {Veitch}, \& {Graff}}]{vitale17}
{Vitale}, S., {Lynch}, R., {Raymond}, V., {et~al.} 2017, \prd, 95, 064053,
  \dodoi{10.1103/PhysRevD.95.064053}

\bibitem[{{Weaver} \& {Woosley}(1993)}]{weaver93b}
{Weaver}, T.~A., \& {Woosley}, S.~E. 1993, \physrep, 227, 65,
  \dodoi{10.1016/0370-1573(93)90058-L}

\bibitem[{{West} {et~al.}(2013){West}, {Heger}, \& {Austin}}]{west13}
{West}, C., {Heger}, A., \& {Austin}, S.~M. 2013, \apj, 769, 2,
  \dodoi{10.1088/0004-637X/769/1/2}

\bibitem[{{Whalen} {et~al.}(2013{\natexlab{a}}){Whalen}, {Fryer}, {Holz},
  {Heger}, {Woosley}, {Stiavelli}, {Even}, \& {Frey}}]{whalen13a}
{Whalen}, D.~J., {Fryer}, C.~L., {Holz}, D.~E., {et~al.} 2013{\natexlab{a}},
  \apjl, 762, L6, \dodoi{10.1088/2041-8205/762/1/L6}

\bibitem[{{Whalen} {et~al.}(2013{\natexlab{b}}){Whalen}, {Even}, {Frey},
  {Smidt}, {Johnson}, {Lovekin}, {Fryer}, {Stiavelli}, {Holz}, {Heger},
  {Woosley}, \& {Hungerford}}]{wahlen13b}
{Whalen}, D.~J., {Even}, W., {Frey}, L.~H., {et~al.} 2013{\natexlab{b}}, \apj,
  777, 110, \dodoi{10.1088/0004-637X/777/2/110}

\bibitem[{{Woosley}(2017)}]{woosley:17}
{Woosley}, S.~E. 2017, \apj, 836, 244, \dodoi{10.3847/1538-4357/836/2/244}

\bibitem[{{Woosley}(2018)}]{woosley:18}
---. 2018, \apj, 863, 105, \dodoi{10.3847/1538-4357/aad044}

\bibitem[{{Woosley}(2019)}]{woosley19}
---. 2019, \apj, 878, 49, \dodoi{10.3847/1538-4357/ab1b41}

\bibitem[{{Woosley} {et~al.}(2007){Woosley}, {Blinnikov}, \&
  {Heger}}]{woosley:07}
{Woosley}, S.~E., {Blinnikov}, S., \& {Heger}, A. 2007, \nat, 450, 390,
  \dodoi{10.1038/nature06333}

\bibitem[{{Woosley} \& {Heger}(2007)}]{woosley07}
{Woosley}, S.~E., \& {Heger}, A. 2007, \physrep, 442, 269,
  \dodoi{10.1016/j.physrep.2007.02.009}

\bibitem[{{Woosley} {et~al.}(2002){Woosley}, {Heger}, \& {Weaver}}]{woosley:02}
{Woosley}, S.~E., {Heger}, A., \& {Weaver}, T.~A. 2002, Rev. Mod. Phys., 74,
  1015, \dodoi{10.1103/RevModPhys.74.1015}

\bibitem[{{Woosley} {et~al.}(1993){Woosley}, {Timmes}, \& {Weaver}}]{woosley93}
{Woosley}, S.~E., {Timmes}, F.~X., \& {Weaver}, T.~A. 1993, Journal of Physics
  G Nuclear Physics, 19, S183, \dodoi{10.1088/0954-3899/19/S/016}

\bibitem[{{Yang} {et~al.}(2019){Yang}, {Bartos}, {Gayathri}, {Ford}, {Haiman},
  {Klimenko}, {Kocsis}, {M{\'a}rka}, {M{\'a}rka}, {McKernan}, \&
  {O'Shaughnessy}}]{yang19}
{Yang}, Y., {Bartos}, I., {Gayathri}, V., {et~al.} 2019, \prl, 123, 181101,
  \dodoi{10.1103/PhysRevLett.123.181101}

\bibitem[{{Yoshida} {et~al.}(2016){Yoshida}, {Umeda}, {Maeda}, \&
  {Ishii}}]{yoshida:16}
{Yoshida}, T., {Umeda}, H., {Maeda}, K., \& {Ishii}, T. 2016, \mnras, 457, 351,
  \dodoi{10.1093/mnras/stv3002}

\bibitem[{{Young} {et~al.}(2008){Young}, {Smartt}, {Mattila}, {Tanvir},
  {Bersier}, {Chambers}, {Kaiser}, \& {Tonry}}]{young08}
{Young}, D.~R., {Smartt}, S.~J., {Mattila}, S., {et~al.} 2008, \aap, 489, 359,
  \dodoi{10.1051/0004-6361:20078662}

\end{thebibliography}

\end{document}